\documentclass[a4paper]{article}
\pdfoutput=1

\usepackage{bbm}
\usepackage{graphicx}
\usepackage{jcappub}
\usepackage{mathtools}
\usepackage{upgreek}
\usepackage{xcolor}
\usepackage{placeins}

\newcommand{\dif}{\mathrm{d}}

\newcommand{\im}{\mathrm{i}}

\newcommand{\ba}{\mathrm{b}}
\newcommand{\dm}{\mathrm{d}}

\newcommand{\revision}[1]{\textcolor{black}{#1}}

\begin{document}

\title{Resummed Kinetic Field Theory: a model of coupled baryonic and dark matter}

\author[a,b,g]{Daniel Geiss,}
\author[c,d,g]{Ivan Kostyuk,}
\author[e,g]{Robert Lilow}
\author[f,g]{and Matthias Bartelmann}

\affiliation[a]{Institut f\"ur Theoretische Physik, Leipzig University, 04013 Leipzig, Germany}
\affiliation[b]{Max Planck Institute for Mathematics in the Sciences, 04013 Leipzig, Germany}
\affiliation[c]{Max Planck Institute for Astrophysics, 85748 Garching, Germany}
\affiliation[d]{Ludwig-Maximilians-Universit\"at, Geschwister-Scholl Platz 1, 80539 Munich, Germany}
\affiliation[e]{Department of Physics, Technion, Haifa 3200003, Israel}
\affiliation[f]{Institut f\"ur Theoretische Physik, Heidelberg University, Philosophenweg 16, 69120 Heidelberg, Germany}
\affiliation[g]{Institut f\"ur Theoretische Astrophysik, Zentrum f\"ur Astronomie, Heidelberg University, Philosophenweg 12, 69120 Heidelberg, Germany}

\emailAdd{daniel.geiss@mis.mpg.de}
\emailAdd{ivkos@mpa-garching.mpg.de}
\emailAdd{rlilow@campus.technion.ac.il}
\emailAdd{bartelmann@uni-heidelberg.de}

\abstract{We present a new analytical description of cosmic structure formation in a mixture of dark and baryonic matter, using the framework of Kinetic Field Theory (KFT) -- a statistical field theory  for classical particle dynamics. So far, KFT has only been able to describe a single type of particles, sufficient to consider structure growth due to the gravitational interactions between dark matter. However, the influence of baryonic gas dynamics becomes increasingly relevant when describing smaller scales. In this paper, we thus demonstrate how to extend the KFT formalism as well as a previously presented resummation scheme towards describing such mixtures of two particle species. Thereby, the gas dynamics of baryons are accounted for using the recently developed model of Mesoscopic Particle Hydrodynamics. Assuming a flat $\Lambda$CDM Universe and a simplified model for the thermal gas evolution, we demonstrate the validity of this approach by computing the linear evolution of the individual and total matter power spectra between the epoch of recombination and today. Our results correctly reproduce the expected behaviour, showing a suppression of both baryonic and dark matter structure growth on scales smaller than the baryonic Jeans length, in good agreement with results from the numerical Boltzmann solver CLASS. Nonlinear corrections within this approach will be investigated in upcoming works.}

\arxivnumber{2007.09484}


\maketitle

\flushbottom
\section{Introduction}

Understanding cosmic structure formation is one of the main goals of current cosmological research. A thorough understanding of it could allow us to derive the current state of the Universe from the fundamental physical properties underlying cosmological inflation models. At present, the most precise predictions for structure growth are obtained from many-body simulations, e.g.~\cite{schaye_eagle_2015,klypin_multidark_2016,springel_first_2018}. However, these methods are numerically very expensive and do not provide deeper insight into the underlying physical processes. The established analytic descriptions of structure formation are based on Eulerian and Lagrangian perturbation theories (see \cite{crocce_renormalized_2006,matarrese_resumming_2007,matsubara_resumming_2008,pietroni_coarse-grained_2012,blas_cosmological_2014,porto_lagrangian-space_2014,senatore_ir-resummed_2015} for a selection of different approaches and \cite{bernardeau_large-scale_2002} for a review). These approaches construct perturbative expansions in density contrast or displacement fields, assuming a single-valued velocity field for cosmic matter. Although they are very successful in describing structure growth on midly nonlinear scales, they are by construction unable to treat the multi-streaming regime of dark matter (DM) dynamics on smaller scales. While it is possible to extend Lagrangian descriptions beyond shell-crossing by explicitly summing over multiple streams, e.g.~\cite{mcdonald_large-scale_2018,pietroni_structure_2018}, we deem a different approach more natural and promising.

Kinetic Field Theory (KFT), an analytic description of structure formation based on classical particle dynamics in phase-space, was developed by Bartelmann et al. in \cite{bartelmann_microscopic_2016,bartelmann_kinetic_2017}, building on the pioneering work of Das and Mazenko \cite{mazenko_fundamental_2010,das_field_2012}. In this theory a partition function is constructed which captures the full Hamiltonian dynamics of the individual particles. Therefore, this analytic framework naturally shares the advantage of numerical simulations of allowing particle streams to cross each other. Previous papers have shown in detail how the evolution of DM power spectra can be calculated down to strongly nonlinear scales in KFT, either by performing a perturbative expansion in orders of the interaction potential, \cite{bartelmann_microscopic_2016,bartelmann_kinetic_2017}, or by finding a suitable averaging procedure for those interactions, \cite{bartelmann_analytic_2017}. A review of the formalism and various applications of KFT can be found in \cite{bartelmann_cosmic_2019}.

However, even a perfectly accurate treatment of collisionless DM dynamics is not sufficient to provide a complete description of structure formation on cosmologically small scales, as baryonic effects play an increasingly important role on scales below the baryonic Jeans length. A comprehensive treatment of baryonic physics, including detailed radiative cooling and astrophysical feedback effects, is currently only accessible via hydrodynamical simulations, see e.g.~\cite{rudd_effects_2008,vogelsberger_introducing_2014,schaye_eagle_2015,mccarthy_bahamas_2017,chisari_impact_2018}. A more modest goal of only taking into account the correct baryonic gas dynamics and potentially some effective treatment of radiative cooling, though, is certainly in the reach of analytical formalisms and should already allow to capture the dominant influence of baryons at sufficiently early times or on sufficiently large scales. While there have been a few works investigating this in Eulerian perturbation theory, e.g.~\cite{nusser_analytic_2000,matarrese_growth_2002,shoji_third-order_2009}, we are aiming at a joint framework capable of accurately describing both the collisionless dynamics of DM as well the gas dynamics of baryons.

To this end, we have developed the model of Mesoscopic Particle Hydrodynamics (MPH) in \cite{viermann_model_2018,geiss_resummed_2019}, allowing to describe baryonic dynamics in KFT. This is achieved by recasting the hydrodynamic equations into equations of motion for effective mesoscopic particles, similar to the numerical method of Smoothed Particle Hydrodynamics \cite{monaghan_smoothed_2005}. A simple perturbative expansion to finite order in the effective MPH interactions, however, was found insufficient to describe hydrodynamics consistently \cite{viermann_model_2018}. In \cite{geiss_resummed_2019} it was shown that this limitation can be overcome by treating MPH in the Resummed KFT (RKFT) framework introduced in \cite{lilow_resummed_2019}, instead. RKFT expands in orders of macroscopic correlation functions, leading to a resummation of infinite subsets of the perturbative expansion in particle interactions.

Until now, KFT has only been used to describe systems containing a single species of particles -- either micro- or mesoscopic ones. The purpose of the present paper is to generalize the formalism of KFT, and specifically RKFT, to systems of two different types of particles, thus allowing us to investigate cosmic structure formation in a mixture of microscopic dark and mesoscopic baryonic matter. In this first application, we test the general feasibility and validity of this approach by analysing the effect of baryonic gas dynamics on the linear evolution of the power spectrum between the epoch of recombination and today, assuming a spatially constant gas temperature that follows the average thermal gas evolution during this time. Despite the applied approximations, our method is able to reproduce the expected suppression of structure growth due to the baryonic pressure and shows good agreement with results from the numerical Boltzmann solver CLASS \cite{blas_cosmic_2011}. A more accurate treatment of the gas temperature as well as nonlinear corrections will be investigated in future papers.

The remainder of this paper is organized as follows. In \autoref{generatingfunctional} we show how to construct a generating functional for a system containing two particle species. Afterwards, in \autoref{freetheory}, we explain how to obtain correlation functions for these mixtures in non-interacting KFT, before deriving the respective resummed correlators in \autoref{resummation}. In \autoref{cosmology} we discuss the specific dynamics of dark and baryonic matter appropriate to investigate cosmic structure formation. Assuming a flat $\Lambda$CDM Universe, we then calculate the linearly evolved density contrast power spectra for both types of matter in \autoref{results} and compare our results to those of a Boltzmann solver. We conclude in \autoref{conclusion} by summarizing the insights gained in this work and providing an outlook on the next steps to be taken.
\section{Generating functional \label{generatingfunctional}}
The basic idea of the KFT formalism is to encapsulate the microscopic dynamics of particles in a generating functional from which the behavior of macroscopic observables can be calculated perturbatively. Therefore, a canonical ensemble consisting of $N$ particles in a volume $V$ is considered. In the following, we apply the approach to a system consisting of two particle species describing baryonic and dark matter. Nevertheless, we want to emphasize the general applicability of this formalism, as it can easily be adapted to other models by changing the respective interaction potentials or even extended to more than two particle species.

\subsection{Two particle species}
The individual particles are described by their phase-space coordinates $\vec{x}^{\alpha}_i \coloneqq (\vec{q}^\alpha_i,\vec{p}^\alpha_i)$, where $\alpha \in \{\mathrm{b},\mathrm{d}\}$ corresponds to the particle species ($\mathrm{b}\,\equiv$ baryons, $\mathrm{d}\,\equiv$ dark matter) and $i=1,...,N^\alpha$ (with $N^\mathrm{b}+N^\mathrm{d}=N$) is labeling the specific particle. The spatial coordinates are denoted by $\vec{q}^\alpha_i$, and $\vec{p}^\alpha_i$ are the respective canonically conjugated momenta. It is convenient to bundle the phase-space coordinates of a single species into the phase-space tensor
\begin{equation}
	\boldsymbol{x}^\alpha \coloneqq \vec{x}^\alpha_i \otimes \vec{e}^\alpha_i,
\end{equation}
where summation over double lower indices is implied and $\{\vec{e}^\alpha_j \}$ denotes the canonical basis in $N^\alpha$ dimensions with entries $(\vec{e}^\alpha_i)_j = \delta_{ij}$. Hence, all microscopic information of the system is contained in the tuple of tensors $\boldsymbol{\vec{x}}\coloneqq(\boldsymbol{x}^\mathrm{b},\boldsymbol{x}^\mathrm{d})$. In this notation a scalar product can be defined as 
\begin{equation}
	\boldsymbol{\vec{a}}\cdot \boldsymbol{\vec{b}} \coloneqq \boldsymbol{a}^\mathrm{b}\cdot\boldsymbol{b}^\mathrm{b} + \boldsymbol{a}^\mathrm{d}\cdot\boldsymbol{b}^\mathrm{d} = \vec{a}^\mathrm{b}_i \cdot \vec{b}^\mathrm{b}_i +  \vec{a}^\mathrm{d}_j \cdot \vec{b}^\mathrm{d}_j.
\end{equation}
The equations of motion are given by Hamilton's equations and take the form
\begin{equation}
	\boldsymbol{E}^\alpha[\boldsymbol{x}^\alpha] = (\partial_t + \boldsymbol{F}) \boldsymbol{x}^\alpha +  \boldsymbol{\nabla}_q V^\alpha = 0  \label{EOM}
\end{equation}
with $\boldsymbol{F}$ describing the linear part of the equation of motion. $V^\alpha(\vec{q}, t)$ denotes an interaction potential which depends on the spatial position as well as time and the particle species $\alpha$, taking the form
\begin{align}
	V^\alpha(\vec{q},t) = \sum_{i=1}^{N^\mathrm{b}} v^{\alpha \mathrm{b}}(\vec{q}-\vec{q}^\mathrm{b}_i,t) + \sum_{j=1}^{N^\mathrm{d}}
	v^{\alpha \mathrm{d}}(\vec{q}-\vec{q}^\mathrm{d}_j,t) \,.
	\label{eq:total_interaction_potential}
\end{align}
The first sum describes interactions of a particle of species $\alpha$ with baryonic and the second with dark matter. Thereby, $v^{\alpha \gamma}$ is the single-particle potential of a $\gamma$-particle as experienced by an $\alpha$-particle.

In the KFT approach \cite{bartelmann_microscopic_2016}, a generating functional is formulated by integrating over all possible trajectories, where a Dirac delta distribution ensures that only phase-space trajectories satisfying the equations of motion \eqref{EOM} contribute. Due to the incomplete information on the initial microscopic state of the system, stochasticity enters the expressions by averaging over the initial conditions according to an initial phase-space probability distribution $P(\vec{\boldsymbol{x}}^{(\mathrm{i})} )$. The corresponding generating functional for a system of two particle species takes the form
\begin{equation}
Z[\vec{\boldsymbol{J}},\vec{\boldsymbol{K}}] \coloneqq \int \mathrm{d}\vec{\boldsymbol{x}}^{(\mathrm{i})} \ P(\vec{\boldsymbol{x}}^{(\mathrm{i})}) \ \int_{\vec{\boldsymbol{x}}^{(\mathrm{i})} } \mathcal{D}\vec{\boldsymbol{x}}(t) \int \mathcal{D}\vec{\boldsymbol{\chi}}(t)\ \exp \bigg\{ \mathrm{i} \int \mathrm{d}t \bigg( \vec{\boldsymbol{\chi}} \cdot \vec{\boldsymbol{E}}[\boldsymbol{x}] + \vec{\boldsymbol{\chi}} \cdot \vec{\boldsymbol{K}} + \vec{\boldsymbol{x}} \cdot \vec{\boldsymbol{J}} \bigg) \bigg\}.
\end{equation}
The appearance of the auxiliary field $\vec{\boldsymbol{\chi}}(t) = (\boldsymbol{\chi}^\mathrm{b}(t),\boldsymbol{\chi}^\mathrm{d}(t))^\intercal$ originates from a Fourier transform of the Dirac delta distribution. Two  source fields $\vec{\boldsymbol{K}}(t) = \big(\boldsymbol{K}^\mathrm{b}(t),\boldsymbol{K}^\mathrm{d}(t)\big)^\intercal$, $\vec{\boldsymbol{J}}(t) = \big(\boldsymbol{J}^\mathrm{b}(t),\boldsymbol{J}^\mathrm{d}(t)\big)^\intercal$ are introduced, allowing the calculation of correlators by taking functional derivatives and setting the source fields to zero afterwards,
\begin{equation}
	\left\langle \vec{\boldsymbol{x}}(t) \otimes \cdots \otimes \vec{\boldsymbol{\chi}}(t')\right\rangle = \frac{\updelta}{\mathrm{i} \updelta \vec{\boldsymbol{J}}(t)} \otimes \cdots \otimes \frac{\updelta}{\mathrm{i} \updelta \vec{\boldsymbol{K}}(t')} \, Z[\vec{\boldsymbol{J}},\vec{\boldsymbol{K}}] \bigg|_{\vec{\boldsymbol{J}},\vec{\boldsymbol{K}}=0}.
\end{equation}

\subsection{Collective fields} \label{Sec: Collective Fields}
Since we are interested in the macroscopic properties of the system instead of the statistics of microscopic fields, we introduce so-called collective fields \cite{bartelmann_microscopic_2016}. A suitable set for such a description is given by the number density $\vec{\Phi}_n \coloneqq \big(\Phi_n^\mathrm{b},\Phi_n^\mathrm{d}\big)^\intercal$, which carries the information on the particles' positions, and the response field $\vec{\Phi}_B \coloneqq \big(\Phi_B^\mathrm{b},\Phi_B^\mathrm{d}\big)^\intercal$, which encodes how the particle momenta are changed by a given interaction potential. In Fourier space they take the form
\begin{align}
\Phi^\alpha_n(\vec{k},t) &\coloneqq \sum_{j=1}^{N} \mathrm{e}^{\mathrm{i} \vec{k}\cdot\vec{q}^\alpha_j(t)}  ,\\
\Phi^\alpha_B(\vec{k},t) &\coloneqq \sum_{j=1}^{N} \mathrm{i} \vec{k} \cdot\vec{\chi}^\alpha_{p_j}(t) \, \mathrm{e}^{-\mathrm{i} \vec{k}\cdot\vec{q}^\alpha_j(t)}
\end{align}
for $\alpha \in \{\ba, \dm\}$. By replacing all occurrences of the fields $\vec{\boldsymbol{x}}$ and $\vec{\boldsymbol{\chi}}$ by functional derivatives with respect to their associated source fields $\vec{\boldsymbol{J}}$ and $\vec{\boldsymbol{K}}$, we define the respective collective-field operators $\hat{\Phi}^\alpha_n$ and $\hat{\Phi}^\alpha_B$. Acting with these on the generating functional of cumulants, i.e.~connected correlators, $W[\vec{\boldsymbol{J}},\vec{\boldsymbol{K}}] \coloneqq \ln Z[\vec{\boldsymbol{J}},\vec{\boldsymbol{K}}]$ and setting the source fields to zero afterwards yields the collective-field cumulants,
\begin{align}
	G^{\alpha_1 \dots \alpha_{l_n} \, \gamma_1 \dots \gamma_{l_B}}_{n \cdots n \, B \cdots B}(1,\dots,l_n,1',\dots,l_B') &= \left\langle \prod_{u=1}^{l_n} \Bigl(\Phi^{\alpha_u}_n(u)\Bigr) \, \prod_{r=1}^{l_B} \Bigl(\Phi^{\gamma_r}_{B}(r')\Bigr) \right\rangle_c
	\label{Interacting collective-field cumulants} \\ 
	&= \prod_{u=1}^{l_n} \Bigl(\hat{\Phi}^{\alpha_u}_n(u)\Bigr) \, \prod_{r=1}^{l_B} \Bigl(\hat{\Phi}^{\gamma_r}_{B}(r')\Bigr) \, W[\vec{\boldsymbol{J}},\vec{\boldsymbol{K}}]\bigg|_{\vec{\boldsymbol{J}},\vec{\boldsymbol{K}}=0} . \nonumber
\end{align}
Here, $\alpha_u, \gamma_r \in \{\mathrm{b},\mathrm{d} \}$ label to which particle species the $u$-th density field and the $r$-th response field correspond, respectively. Pure $\vec{\Phi}_n$-cumulants describe the connected $l_n$-point correlations of the number density at times $t_1,...t_{l_n}$. Cumulants also involving $\vec{\Phi}_B$-fields, on the other hand, describe the response of those density cumulants to perturbations of the system at times $t_{1'} ,\dots , t_{l'_B}$ caused by a given interaction potential. Note that pure $\vec{\Phi}_B$-field cumulants vanish, as they do not contain any density fields whose response they would characterize.

With the collective fields at hand the generating functional can be rewritten as
\begin{equation}
Z[\vec{\boldsymbol{J}},\vec{\boldsymbol{K}}] = \mathrm{e}^{\mathrm{i} \hat{S}_\mathrm{I} } Z_0[\vec{\boldsymbol{J}},\vec{\boldsymbol{K}}] \label{Basics: Full Gen. Func. for Class. Fields}
\end{equation}
with $Z_0$ describing the non-interacting system and $\hat{S}_\mathrm{I}$ encapsulating the interactions between particles. The free generating functional $Z_0$ is given by
\begin{equation}
Z_0[\vec{\boldsymbol{J}},\vec{\boldsymbol{K}}] \coloneqq \int \mathrm{d}\vec{\boldsymbol{x}}^{(\mathrm{i})}\ P(\vec{\boldsymbol{x}}^{(\mathrm{i})}) \mathrm{e}^{\mathrm{i} \! \int \! \mathrm{d}t \, \vec{\boldsymbol{J}} \cdot \vec{\boldsymbol{x}}^\mathrm{lin} }  , \label{Free Generating Functional}
\end{equation}
where $\vec{\boldsymbol{x}}^\mathrm{lin}(t) \coloneqq \big(\boldsymbol{x}^\mathrm{lin,b}(t), \boldsymbol{x}^\mathrm{lin,d}(t)\big)^\intercal$ is the solution to the linear, non-interacting part of the equations of motion, augmented by the source term $\vec{\boldsymbol{K}}$, given by
\begin{equation}
	\boldsymbol{x}^{\mathrm{lin},\alpha}(t) \coloneqq \boldsymbol{\mathcal{G}}^{\mathrm{R}\alpha}(t,t_\mathrm{i}) \, \boldsymbol{x}^{\alpha(\mathrm{i})} - \int_{t_\mathrm{i}}^\infty \mathrm{d}t'\ \boldsymbol{\mathcal{G}}^{\mathrm{R}\alpha}(t,t') \, \boldsymbol{K}^\alpha(t').
\end{equation}
Here, $t_\mathrm{i}$ is the initial time and $\boldsymbol{\mathcal{G}}^{\mathrm{R} \alpha}$ denotes the retarded Green's function of the free equations of motion for $\alpha$-particles, which we will refer to as the microscopic propagator. In general, it can be expressed as \cite{bartelmann_microscopic_2016}
\begin{align}
	\boldsymbol{\mathcal{G}}^{\mathrm{R}\alpha}(t,t') &\coloneqq  \mathcal{G}^\mathrm{R}(t,t') \otimes \mathbbm{1}_{N^\alpha}, \\ \mathcal{G}^\mathrm{R}(t,t') &\coloneqq
	\begin{pmatrix}
		g_{qq}(t,t') \, \mathbbm{1}_3 \quad & g_{qp}(t,t') \, \mathbbm{1}_3 \\
		g_{pq}(t,t') \, \mathbbm{1}_3 \quad & g_{pp}(t,t') \, \mathbbm{1}_3 \\
	\end{pmatrix}
	\propto \uptheta(t-t') ,
	\label{MPH: propagator}
\end{align}
with $\mathbbm{1}_d$ denoting the $d \times d$ identity matrix.
Note that we can use the same single-particle propagator $\mathcal{G}^\mathrm{R}$ for both particle species at this point only when imposing that both types of particles have the same mass. Since we are only interested in macroscopic fields and the particle picture is only a feature of the mathematical formalism this indeed poses a valid choice of parameters. To control different (average) mass densities we simply adapt the ratio of the mean number densities $\bar{n}^\alpha$.

The interaction part of the action takes the form
\begin{align}
	S_\mathrm{I}[\vec{\boldsymbol{x}},\vec{\boldsymbol{\chi}}] 
	= &- \int \mathrm{d}1\ \vec{\Phi}_n(1)\underline{v}(1)\vec{\Phi}_B(-1)
	\label{Interacting part of microscopic action} \\
	\eqqcolon &-\vec{\Phi}_n \cdot \underline{v} \cdot \vec{\Phi}_B, \nonumber
\end{align}
where we introduced the abbreviations $f(\pm m) \coloneqq f(\pm\vec{k}_m,t_m)$ and $\mathrm{d}m\coloneqq\frac{\mathrm{d}^3k_m}{(2\pi)^3} \mathrm{d}t_m$, and $\underline{v}$ denotes the interaction potential matrix given by
\begin{equation}
	\underline{v} \coloneqq 
	\begin{pmatrix}
		v^\mathrm{bb} & v^\mathrm{bd} \\
		v^\mathrm{db} & v^\mathrm{dd}
	\end{pmatrix}.
	\label{Interaction potential matrix}
\end{equation}
Furthermore, the hat above $\hat{S}_\mathrm{I}$ in \eqref{Basics: Full Gen. Func. for Class. Fields} indicates that all fields are replaced by appropriate functional derivatives with respect to the corresponding source fields.

\subsection{Initial distribution\label{Sec: Init Distr}}
The initial distribution $P(\boldsymbol{\vec{x}}^\mathrm{(i)})$ of the microscopic particles is obtained by a sampling of Gaussian distributed initial macroscopic density and momentum fields \cite{bartelmann_microscopic_2016}. Furthermore, explicit constraints for the covariance, which describes the initial density and momentum auto- as well as cross-correlations, can be formulated by imposing statistical homogeneity and isotropy, thus reflecting the large-scale nature of the early Universe. The explicit calculation of the initial distribution for a system of two particle species can be performed completely analogously to the single-species case. A schematic derivation is given in \autoref{App: Polynomial}, for further detail we refer to \cite{bartelmann_microscopic_2016}. The result for the initial phase-space distribution is found to be
\begin{align}
P(\vec{\boldsymbol{x}}^{(\mathrm{i})}) =
\frac{V^{-N}}{\sqrt{(2\pi)^{3N} \det \boldsymbol{C}_{pp}}}
\hat{\mathcal{C}} \bigg( \frac{\partial}{\mathrm{i} \partial \vec{\boldsymbol{p}}^{(\mathrm{i})}} \bigg) \exp \bigg\{ -\frac{1}{2} \vec{\boldsymbol{p}}^{(\mathrm{i})\intercal} \boldsymbol{C}^{-1}_{pp} \vec{\boldsymbol{p}}^{(\mathrm{i})}  \bigg\}  \label{MPH: Initial Distribution}
\end{align}
with the combined initial momentum tensor and the corresponding covariance matrix
\begin{align}
\vec{\boldsymbol{p}}^{(\mathrm{i})} \coloneqq
\begin{pmatrix}
\boldsymbol{p}^{\mathrm{b}(\mathrm{i})}  \\  \boldsymbol{p}^{\mathrm{d}(\mathrm{i})}
\end{pmatrix},
\qquad
\boldsymbol{C}_{pp} \coloneqq 
\begin{pmatrix}
C_{pp}^{\mathrm{b}\mathrm{b}} & C_{pp}^{\mathrm{b}\mathrm{d}} \\
C_{pp}^{\mathrm{d}\mathrm{b}} &
C_{pp}^{\mathrm{d}\mathrm{d}}
\end{pmatrix}.
\end{align}
Furthermore, a polynomial operator $\hat{\mathcal{C}} \big( \frac{\partial}{\mathrm{i} \partial \vec{\boldsymbol{p}}^{(\mathrm{i})}} \big)$ is introduced which imposes the initial density as well as density-momentum correlations and is defined in in equation \eqref{polynomial def} of \autoref{App: Polynomial}.

\revision{Since we fix the initial conditions at a sufficiently early time for the linearised continuity equation to hold, we can use this to relate the initial density and momentum fields. Combined with the assumption of Gaussianity this implies that the initial phase-space distribution \eqref{MPH: Initial Distribution} is completely determined by specifying the initial density contrast auto- and cross-power spectra $P_\delta^{\ba \ba \, (\mathrm{i})}$, $P_\delta^{\dm \dm \, (\mathrm{i})}$ and $P_\delta^{\ba \dm \, (\mathrm{i})}$ of dark and baryonic matter.}
\section{Free theory \label{freetheory}}
Later on, we perform a resummation procedure which formally integrates out all microscopic degrees of freedom, leading to a theory formulated in terms of macroscopic quantities only. Since this approach requires knowledge of the free collective-field cumulants, we explain the procedure to calculate these. The calculation is very technical and mostly analogous to the case of only one particle species as presented in \cite{fabis_kinetic_2018}. While giving a few more details, with focus on the differences to the one-species case, in \autoref{App: Free Gen}, we will only state the key results necessary to understand the conceptual idea of the procedure here.

\subsection{Free generating functional}
\label{Sec: Free GenFunc}
The full free generating functional is obtained by inserting \eqref{MPH: Initial Distribution} into \eqref{Free Generating Functional}. After several manipulations, which amongst others includes a decomposition of the momentum correlation matrix into cross-correlations between different particles, $\tilde{C}_{pp}$, and self-correlations (see \autoref{App: Free Gen}), one finds that the free generating functional is of the form
\begin{align}
Z_{0}[\vec{\boldsymbol{J}},\vec{\boldsymbol{K}}] = \int \mathrm{d}\vec{\boldsymbol{x}}^{(\mathrm{i})}\  &\hat{\mathcal{C}}_\mathrm{tot}\bigg( \frac{\updelta}{\mathrm{i} \updelta \vec{\boldsymbol{K}}_p(t_\mathrm{i})} \bigg) \frac{P_{\sigma^{\mathrm{b}}_p}(\boldsymbol{p}^{\mathrm{b}(\mathrm{i})})}{V^{N^\mathrm{b}}} \frac{P_{\sigma^{\mathrm{d}}_p}(\boldsymbol{p}^{\mathrm{b}(\mathrm{i})})}{V^{N^\mathrm{d}}}    \label{MPH: Coupled Free Gen Func}\\ 
&\times \exp\bigg\{ \mathrm{i} \bigg( \vec{\boldsymbol{\mathcal{J}}}_q \vec{\boldsymbol{q}}^{(\mathrm{i})} + \vec{\boldsymbol{\mathcal{J}}}_p \vec{\boldsymbol{p}}^{(\mathrm{i})} -S_K[\vec{\boldsymbol{J}},\vec{\boldsymbol{K}}] \bigg) \bigg\}  \nonumber,
\end{align}
where $P_{\sigma^{\alpha}_p}$ denote uncorrelated Gaussian distributions defined in equation \eqref{Gaussian distr} of \autoref{App: Free Gen} and
\begin{align}
\hat{\mathcal{C}}_\mathrm{tot} \bigg( \frac{\updelta}{\mathrm{i} \updelta \vec{\boldsymbol{K}}_p(t_\mathrm{i})} \bigg) \coloneqq \hat{\mathcal{C}} \bigg( \frac{\updelta}{\mathrm{i} \updelta \vec{\boldsymbol{K}}_p(t_\mathrm{i})} \bigg) \exp \bigg\{ -\frac{1}{2} \bigg( \frac{\updelta}{\mathrm{i} \updelta \vec{\boldsymbol{K}}_p(t_\mathrm{i})} \bigg)^\intercal \tilde{\boldsymbol{C}}_{pp} \bigg( \frac{\updelta}{\mathrm{i} \updelta \vec{\boldsymbol{K}}_p(t_\mathrm{i})} \bigg) \bigg\}
\end{align}
inherits all information on correlations between different particles. In the case of vanishing initial cross-correlations, corresponding to $\hat{\mathcal{C}}_\mathrm{tot} \rightarrow 1$, the free generating functional factorizes into single-particle contributions. Then, the free evolution of the whole system is fully determined by the generating functional of a single particle\footnote{Since we assume the same mass for both particle species their free evolution is equivalent. If we considered particle species of different masses, we would have to take the different single-particle generating functionals of both species into account.}. If we turn on cross-correlations, the particles become statistically related such that the generating functional can no longer be separated into one-particle contributions. Rather, the factorization can now be performed in terms of clusters of correlated particles. A diagrammatic representaion for these based on the Mayer cluster expansion \cite{mayer_molecular_1941} was developed in \cite{fabis_kinetic_2018}, allowing for a systematic computation of the free collective-field cumulants.

\subsection{Free cumulants}\label{Section: Free Cumulants}
Due to the factorization of the free generating functional, the free cumulants can be formulated as a sum over contributions from clusters of $\ell^\mathrm{b}$ correlated baryonic and $\ell^\mathrm{d}$ correlated dark-matter particles,
\begin{align}
    G^{(0) \, \alpha_1 \dots \alpha_{l_n} \, \gamma_1 \dots \gamma_{l_B}}_{\hphantom{(0)} \, n \cdots n \, B \cdots B}(1,\dots,l_n,1',\dots,l_B') &= \prod_{u=1}^{l_n} \Bigl(\hat{\Phi}^{\alpha_u}_n(u)\Bigr) \, \prod_{r=1}^{l_B} \Bigl(\hat{\Phi}^{\gamma_r}_{B}(r')\Bigr) \, W_0 [\vec{\boldsymbol{J}},\vec{\boldsymbol{K}}] \biggr|_{\vec{\boldsymbol{J}},\vec{\boldsymbol{K}}=0}   \label{MPH: Coupled Free Cumulants Equation} \\
    &\eqqcolon \sum_{\substack{\ell^\mathrm{b},\ell^\mathrm{d} = 0 \\ \ell^\mathrm{b}+\ell^\mathrm{d} \geq 1}}^\infty G^{(\ell^\mathrm{b},\ell^\mathrm{d}) \, \alpha_1 \dots \alpha_{l_n} \, \gamma_1 \dots \gamma_{l_B}}_{0 \hphantom{(,\ell^\mathrm{d})} \;\, n \cdots n \, B \cdots B} (1,\dots,l_n,1',\dots,l_B') , \nonumber
\end{align}
where $W_0[\vec{\boldsymbol{J}},\vec{\boldsymbol{K}}] \coloneqq \ln Z_0[\vec{\boldsymbol{J}},\vec{\boldsymbol{K}}]$ and $\alpha_u, \gamma_r \in \{\mathrm{b},\mathrm{d} \}$. The explicit calculation of the free $(\ell^\mathrm{b},\ell^\mathrm{d})$-particle cumulants appearing in the second line of \eqref{MPH: Coupled Free Cumulants Equation} is quite involved and described in more detail in \autoref{App: Free Cumulants}. From a rigorous analysis, as it was done in \cite{fabis_kinetic_2018}, one can infer certain rules for which terms contribute to the sum in \eqref{MPH: Coupled Free Cumulants Equation}. For our case of two particle species these rules are formulated in \autoref{homogeneity rule} and \autoref{causality rule} of \autoref{App: Free Cumulants}. One central finding is that the sums over $\ell^\alpha$ in \eqref{MPH: Coupled Free Cumulants Equation} truncate at the number of the respective density fields $\Phi_n^\alpha$ appearing in the cumulant, such that only a small finite number of terms actually needs to be computed. This is because one needs at least an $\ell^\alpha$-point $\Phi_n^\alpha$-density cumulant to describe the mutual correlations between $\ell^\alpha$ different particles. Moreover, as already stated earlier, pure $\vec{\Phi}_B$-field cumulants vanish. 

In the following, we calculate the power spectra from the resummed theory at linear level only. For this, we only need the free 2-point cumulants. Assuming a statistically isotropic and homogenous system, suitable for a cosmological application, the free 2-point cumulants expanded to first order in the initial auto- and cross-spectra $P^{\alpha \gamma \, (\mathrm{i})}_\delta$ of baryonic and dark matter, disregarding shot noise, take the form 
\begin{align}
G^{(0)}_{\vec{B}\vec{B}}(1,2) &= 0,  \label{2-pt Cumulants 1}\\
G^{(0)}_{\vec{n}\vec{B}}(1,2) = G^{(0)}_{\vec{B}\vec{n}}(2,1) &\approx -i(2\pi)^3 \updelta_\textsc{d}(\vec{k}_1+\vec{k}_2) \, \vec{k}_1^2 \, g_{qp}(t_1,t_2)  
\begin{pmatrix}
\bar{n}^\mathrm{b} & 0 \\
0 & \bar{n}^\mathrm{d} \\
\end{pmatrix}, \label{2-pt Cumulants 2}\\
G^{(0)}_{\vec{n}\vec{n}}(1,2) &\approx (2\pi)^3 \updelta_\textsc{d}(\vec{k}_1+\vec{k}_2) \big(1+g_{qp}(t_1,0)\big)\big(1+g_{qp}(t_2,0)\big) \label{2-pt Cumulants 3}\\
&\quad\;\times\begin{pmatrix}
\bar{n}^{\mathrm{b}}\bar{n}^{\mathrm{b}} P^{\mathrm{bb} \, (\mathrm{i})}_\delta\revision{(k_1)} \quad & \bar{n}^\mathrm{b}\bar{n}^\mathrm{d} P^{\mathrm{bd} \, (\mathrm{i})}_\delta\revision{(k_1)}\\
\bar{n}^\mathrm{d}\bar{n}^\mathrm{b} P^{\mathrm{db} \, (\mathrm{i})}_\delta\revision{(k_1)}\quad & \bar{n}^{\mathrm{d}}\bar{n}^{\mathrm{d}} P^{\mathrm{dd} \, (\mathrm{i})}_\delta\revision{(k_1)} \\
\end{pmatrix}  ,\nonumber
\end{align} 
conveniently written in terms of $2 \times 2$ matrices. Note that the off-diagonal components of $G^{(0)}_{\vec{n}\vec{B}}$, i.e.~$G^{(0) \, \mathrm{d}\mathrm{b}}_{\hphantom{(0) \,} nB}$ and $G^{(0) \, \mathrm{b}\mathrm{d}}_{\hphantom{(0) \,} nB}$, vanish since the field $\Phi^\alpha_B$ only describes the response of the $\alpha$-particle density $\Phi^\alpha_n$ to perturbations but not the response of the density of the other particle species.
\section{Resummation \label{resummation}}
For the description of structure formation in pure collisionless DM, the microscopic perturbative approach to KFT, corresponding to an expansion of the exponential in \eqref{Basics: Full Gen. Func. for Class. Fields} in orders of the interaction operator $\hat{S}_\mathrm{I}$, has proven itself very successful, see e.g.~\cite{bartelmann_microscopic_2016,bartelmann_kinetic_2017}. However, when this approach was applied to baryonic matter formulated in terms of the MPH approach in \cite{viermann_model_2018}, indications were found that any finite-order expansion in $\hat{S}_\mathrm{I}$ is probably insufficient to treat fluid dynamics consistently. Hence, for an analysis of the cosmic structure formation including baryonic matter another approach becomes necessary.

In this context, a reformulation of the original KFT approach, dubbed Resummed KFT (RKFT), was proposed in \cite{lilow_resummed_2019} and later successfully applied to MPH in \cite{geiss_resummed_2019}. In this reformulation the microscopic degrees of freedom are formally integrated out such that the generating functional is formulated in terms of macroscopic fields only. As a result, even the lowest-order perturbative calculation within RKFT involves the resummation of an infinite subset of terms appearing in the microscopic perturbative expansion in orders of $\hat{S}_\mathrm{I}$. Crucially, this reformulation is exact and thus preserves all information on the micro- and mesoscopic particle dynamics. This is possible because the freely-evolving system is exactly solvable and the interacting part of the action \eqref{Interacting part of microscopic action} depends on the microscopic fields only implicitly via the collective fields.

The generalisation of the RKFT formalism to systems of two particle species is quite straightforward and leaves the overall structure of the associated macroscopic perturbation theory completely unchanged. The only difference to the original derivation in \cite{lilow_resummed_2019} is that all macroscopic fields and cumulants acquire the same particle-species substructure as the collective fields and cumulants described in \autoref{Sec: Collective Fields} and \autoref{Section: Free Cumulants}. In the following, we will thus only summarize the derivation of two-species RKFT and highlight the main differences to the single-species case. We further choose to work with the number density $n$ as the central macroscopic field instead of the Klimontovich phase-space density $f$ used in \cite{lilow_resummed_2019}, as we are not interested in computing cumulants involving the momentum density or any other momentum moments here.

\subsection{Macroscopic generating functional}
In RKFT, the generating functional \eqref{Basics: Full Gen. Func. for Class. Fields} is reformulated in terms of the macroscopic number density field $\vec{n} \coloneqq (n^\mathrm{b}, n^\mathrm{d})$ of the two particle species and a macroscopic auxiliary field $\vec{\beta} \coloneqq (\beta^\mathrm{b}, \beta^\mathrm{d})$. Combining these two fields into the combined macroscopic field $\phi \coloneqq (\vec{n},\vec{\beta})$ and introducing an associated macroscopic source field $M \coloneqq (\vec{M}_n,\vec{M}_\beta)$, the generating functional of macroscopic-field correlators is given by
\begin{equation}
Z_\phi[M] \coloneqq \int \mathcal{D}\phi \, \exp\biggl\{ \mathrm{i} S_\Delta[\phi] + \mathrm{i} S_\mathcal{V}[\phi] + \int \mathrm{d}1 \, M^\top\!(1) \; \phi(-1) \biggr\} .
\label{RKFT Z}
\end{equation}
Here, $S_\Delta$ and $S_\mathcal{V}$ denote the propagator and vertex parts of the macroscopic action, respectively,
\begin{align}
\mathrm{i} S_\Delta[\phi] &\coloneqq -\frac{1}{2} \int \mathrm{d}1 \int \mathrm{d}2 \; \phi^\top\!(-1) \; \Delta^{-1}(1,2) \; \phi(-2) , \\
\mathrm{i} S_\mathcal{V}[\phi] &\coloneqq \sum_{\substack{l_\beta,l_n=0 \\ l_\beta+l_n\neq 2} }^{\infty} \frac{1}{l_\beta! \, l_n! } \prod_{u=1}^{l_\beta} \left(\sum_{\alpha_u \in \{\mathrm{b},\mathrm{d}\}} \int \mathrm{d}u \, \beta^{\alpha_u}(-u) \right) 
\prod_{r=1}^{l_n} \left(\sum_{\gamma_r \in \{\mathrm{b},\mathrm{d}\}} \int \mathrm{d}r' \, n^{\gamma_r}(-r') \right)  \label{vertex part of the macroscopic action} \\
& \qquad \qquad \qquad \quad \; \times \mathcal{V}^{\alpha_1 \cdots \alpha_{l_\beta} \gamma_1 \cdots \gamma_{l_n}}_{\beta\cdots\beta \, n\cdots n}(1,\dots,l_\beta,1',\dots,l_n') . \nonumber
\end{align}
The inverse propagator $\Delta^{-1}$ and the $(l_\beta+l_n)$-point vertices $\mathcal{V}^{\alpha_1 \cdots \alpha_{l_\beta} \gamma_1 \cdots \gamma_{l_n}}_{\beta\cdots\beta \, n\cdots n}$ can be expressed in terms of the free collective-field cumulants $G^{(0) \, \alpha_1 \cdots \alpha_{l_\beta} \gamma_1 \cdots \gamma_{l_n}}_{\hphantom{(0)}\, n \dotsm n B \dotsm B}$ defined in \eqref{MPH: Coupled Free Cumulants Equation} and the interaction potential matrix $\underline{v}$ given in \eqref{Interaction potential matrix},
\begin{align}
\Delta^{-1}(1,2) &=
\begin{pmatrix}
\Delta_{\vec{n} \vec{n}}(1,2) \;\;&\;\; \Delta_{\vec{n} \vec{\beta}}(1,2) \\[0.5em]
\Delta_{\vec{\beta} \vec{n}}(1,2) \;\;&\;\; \Delta_{\vec{\beta} \vec{\beta}}(1,2) \\
\end{pmatrix}^{-1}
\label{inverse propagator} \\[0.5\baselineskip] &=
\begin{pmatrix}
\underline{v}(1) \, G^{(0)}_{\vec{B} \vec{B}}(1,2) \, \underline{v}(2) &\; \mathrm{i} \, \mathcal{I}(1,2) \mathbbm{1}_2 - \underline{v}(1) \, G^{(0)}_{\vec{B} \vec{n}} (1,2) \\[0.5em]
\mathrm{i} \, \mathcal{I}(1,2) \mathbbm{1}_2 - G^{(0)}_{\vec{n} \vec{B}} (1,2) \, \underline{v}(2) & G^{(0)}_{\vec{n} \vec{n}} (1,2) \\
\end{pmatrix} \!,
\nonumber \\[0.5\baselineskip]
\mathcal{V}^{\alpha_1 \cdots \alpha_{l_\beta} \gamma_1 \cdots \gamma_{l_n}}_{\beta\cdots\beta \, n\cdots n}(1,\dots,l_\beta,1',\dots,l_n') &= \mathrm{i}^{l_\beta} \, (-\mathrm{i})^{l_n} \, \prod_{r=1}^{l_n} \left(\sum_{\epsilon_r \in \{\mathrm{b},\mathrm{d}\}} v^{\epsilon_r \gamma_r}(r')\right)
\label{macroscopic_vertices} \\
&\quad\times G^{(0) \,\alpha_1 \cdots \alpha_{l_\beta} \epsilon_1 \cdots \epsilon_{l_n}}_{\hphantom{(0)}\, n \cdots n \, B \cdots B}(1,\dots,l_\beta,1',\dots,l_n') , \nonumber
\end{align}
with the identity 2-point function
\begin{equation}
\mathcal{I}(1,2)\coloneqq (2\pi)^3 \, \delta_\mathrm{D}(\vec{k}_1+\vec{k}_2) \, \delta_\mathrm{D}(t_1-t_2) .
\end{equation}

Note that in contrast to the case of a single particle species, we have to take into account that the free 2-point cumulants and hence all components of the propagator become $2\times2$ matrices, rendering the whole macroscopic propagator $\Delta$ a $4 \times 4$ matrix. Similarly, the $(l_\beta + l_n)$-point vertices have $2^{l_\beta + l_n}$ components. For 2-point quantities we will usually use the compact matrix notation, while we will write out quantities with more arguments in components to avoid notational ambiguity.

To obtain the interacting cumulants of the macroscopic fields, one has to take the desired number of functional derivatives of the cumulant-generating functional $W_\phi[M]\coloneqq \ln Z_\phi[M]$ with respect to the source field $M$ evaluated at $M=0$, 
\begin{align}
G^{\alpha_1 \cdots \alpha_{l_n} \gamma_1 \cdots \gamma_{l_\beta}}_{n\cdots n \, \beta\cdots \beta}(1,\dots,l_n,1',\dots,l_\beta') = \prod_{u=1}^{l_n} \biggl(\frac{\updelta}{\mathrm{i}\updelta M^{\alpha_u}_n(u)} \biggr) \prod_{r=1}^{l_\beta} \biggl(\frac{\updelta}{\mathrm{i}\updelta M^{\gamma_r}_\beta(r')} \biggr) \, W_\phi[M] \bigg|_{M=0}.
\label{macroscopic cumulants}
\end{align}

\subsection{Macroscopic perturbation theory \label{sec:resummation:perturbation_theory}}
In a similar fashion to how we obtained the interaction operator in \eqref{Basics: Full Gen. Func. for Class. Fields}, we can pull out the vertex part of the macroscopic action in front of the path integral by replacing all macroscopic fields appearing in \eqref{vertex part of the macroscopic action} with functional derivatives with respect to the corresponding source fields, $\hat{S}_\mathcal{V} \coloneqq S_\mathcal{V}\bigl[\frac{\updelta}{\mathrm{i} \updelta M}\bigr]$. The remaining Gaussian path integral can then be performed exactly, leaving us with
\begin{equation}
Z_\phi[M] = \mathrm{e}^{\mathrm{i}\hat{S}_\mathcal{V}} \, \exp\biggl\{  -\frac{1}{2} \int \mathrm{d}1 \int \mathrm{d}2 \; M^\top\!(-1) \; \Delta(1,2) \; M(-2) \biggr\} .
\label{macroscopic generating functional with vertex operator} 
\end{equation}
The macroscopic perturbation theory of RKFT is constructed by expanding the first exponential in \eqref{macroscopic generating functional with vertex operator} in orders of the vertices. This allows to express the perturbative contributions to any macroscopic cumulant \eqref{macroscopic cumulants} as combinations of propagators and vertices. A representation of these contributions in terms of Feynman diagrams is given in \cite{lilow_resummed_2019}.

To describe cosmic structure formation, we are especially interested in the density contrast \revision{auto- and cross-power spectra $P_\delta^{\alpha \gamma}$ of baryonic and dark matter as well as the total matter spectrum $P_\delta^\mathrm{(tot)}$}, which can be calculated from the density 2-point number density cumulant $G_{nn}$ according to
\begin{align}
\revision{P_\delta^{\alpha \gamma}(k_1,t_1)} &\revision{= \frac{1}{\bar{n}^\alpha \, \bar{n}^\gamma} \int \frac{\mathrm{d}^3 k_2}{(2\pi)^3} \int \mathrm{d} t_2 \, \updelta_\textsc{d}(t_1-t_2) \, G^\mathrm{\alpha \gamma}_{nn}(1,2) \,,} \label{individual_particle_species_power_spectrum} \\ 
P_\delta^\mathrm{(tot)}(k_1,t_1) &= \frac{1}{(\bar{n}^\mathrm{b} + \bar{n}^\mathrm{d})^2} \int \frac{\mathrm{d}^3 k_2}{(2\pi)^3} \int \mathrm{d} t_2 \, \updelta_\textsc{d}(t_1-t_2) \, \Big( G^\mathrm{bb}_{nn}(1,2) + G^\mathrm{bd}_{nn}(1,2) \label{total_matter_power_spectrum} \\
&\hphantom{= \frac{1}{(\bar{n}^\mathrm{b} + \bar{n}^\mathrm{d})^2} \int \frac{\mathrm{d}^3 k_2}{(2\pi)^3} \int \mathrm{d} t_2 \, \updelta_\textsc{d}(t_1-t_2) \,} + G^\mathrm{db}_{nn}(1,2) + G^\mathrm{dd}_{nn}(1,2) \Big) \,. \nonumber 
\end{align}
In this work, we restrict our analysis to the leading-order (also denoted as tree-level) result, which is obtained from the $\vec{n}\vec{n}$-component of the propagator, $G^{(\mathrm{tree})}_{\vec{n}\vec{n}}=\Delta_{\vec{n}\vec{n}}$, as this suffices to describe the linear evolution of structures. For the calculation of the propagator via \eqref{inverse propagator}, a combined matrix and functional inversion is required,
\begin{equation}
\int \mathrm{d}\bar{1} \; \Delta(1,\bar{1}) \; \Delta^{-1}(-\bar{1},2) = \mathcal{I}(1,2) \, \mathbbm{1}_4 .
\end{equation}
Performing the matrix inversion yields
\begin{align}
G^{(\mathrm{tree})}_{\vec{n}\vec{n}}\revision{(1,2)} = \Delta_{\vec{n}\vec{n}}(1,2) = G^{(0)}_{\vec{n}\vec{n}}(1, 2) &+ \int \mathrm{d}\bar{1} \; \tilde{\Delta}_\textsc{r}(1, -\bar{1}) \, G^{(0)}_{\vec{n}\vec{n}}(\bar{1}, 2) + \int \mathrm{d}\bar{2} \; G^{(0)}_{\vec{n}\vec{n}}(1, \bar{2}) \, \tilde{\Delta}_\textsc{a}(-\bar{2}, 2) \nonumber \\
&+ \int \mathrm{d}\bar{1} \int \mathrm{d}\bar{2} \; \tilde{\Delta}_\textsc{r}(1, -\bar{1}) \, G^{(0)}_{\vec{n}\vec{n}}(\bar{1}, \bar{2}) \, \tilde{\Delta}_\textsc{a}(-\bar{2}, 2) ,
\label{nn-component of the propagator}
\end{align}
with $\tilde{\Delta}_\textsc{r}(1,2) = \tilde{\Delta}_\textsc{a}(2,1)$ being the so-called retarded and advanced macroscopic propagators, respectively, describing the linear response of the number density at time $t_1$ to perturbations of the system at time $t_2$. They are defined as the solution of the matrix integral equation
\begin{equation}
\tilde{\Delta}_\textsc{r}(1, 2) = - \mathrm{i} G^{(0)}_{\vec{n} \vec{B}}(1, 2) \, \underline{v}(2) - \int \mathrm{d}\bar{1} \; \mathrm{i} G^{(0)}_{\vec{n} \vec{B}}(1, \bar{1}) \, \underline{v}(\bar{1}) \, \tilde{\Delta}_\textsc{r}(-\bar{1}, 2) ,
\label{integral equation for retarded propagator}
\end{equation}
where the matrix product under the integral captures the mutual interactions between the two particle species. Formally, this equation can be solved by iteratively inserting the whole right-hand-side into the $\tilde{\Delta}_\textsc{r}$ under the integral, which demonstrates that the tree-level RKFT result already contains contributions of arbitrarily high order in the interaction potential.

It was shown in \cite{lilow_resummed_2019} that in the large-scale limit of cosmic structure formation in purely gravitationally interacting DM, an exact analytic solution to this equation exists that precisely recovers the usual linear growth factor. However, for a solution valid on all scales which also takes into account the pressure effects of baryons, a numerical treatment is necessary. For this purpose, we first exploit that in a statistically homogeneous situation, such as cosmic structure formation, the integral over $\vec{k}_{\bar{1}}$ can be performed trivially since the cumulant $\revision{G^{(0)}_{\vec{n} \vec{B}}}(1,\bar{1})$ is proportional to $\delta_\mathrm{D}(\vec{k}_1 + \vec{k}_{\bar{1}})$. The remaining integral over $t_{\bar{1}}$ can then be approximated by a direct sum over a discrete set of $N_t$ time steps. Overall, this transforms \eqref{integral equation for retarded propagator} into a linear $2N_t \times 2N_t$ matrix equation. Solving this matrix equation is numerically inexpensive and works analogously to the one-species case described in more detail in \cite{lilow_resummed_2019}.
\section{Cosmological setting \label{cosmology}}
In the cosmological context, the two particle species are associated with dark and baryonic matter. While DM only interacts via gravity, the interactions between baryonic particles cause additional pressure effects. For a particle description of baryonic matter we make use of the model of Mesoscopic Particle Hydrodynamics (MPH) which was presented in \cite{viermann_model_2018,geiss_resummed_2019}. In this model so-called mesoscopic particles are introduced, representing fluid elements on a mesoscopic scale $\sigma$ which is chosen such that the associated microscopic particles are in local equilibrium but which is much smaller than the scale of interest. According to the local equilibrium hypothesis, it is possible to define thermodynamic quantities such as pressure on this mesoscopic scale. In what follows, we will summarize the derivation of the meso- and microscopic Green's functions and interaction potentials in an expanding space-time, detailed in \cite{geiss_resummed_2019}, and discuss the cosmological evolution of the baryonic gas pressure.

\subsection{Particle dynamics in an expanding space-time}
The equations of motion governing the dynamics of the mesoscopic fluid are the Euler equations describing mass, momentum and energy conservation. In general, a mesoscopic particle is accordingly characterised by three properties: its position, its momentum and its internal energy or enthalpy. However, for the calculations in this paper we assume a spatially constant gas temperature whose evolution will be fixed externally. As shown in \cite{geiss_resummed_2019}, this renders the energy conservation equation obsolete and allows to describe a mesoscopic particle just by its position and momentum. Furthermore, in KFT the mass conservation or continuity equation is fulfilled by construction since the number of particles is conserved. Hence, we are only left with the momentum conservation equation
\begin{equation}
    \rho \frac{\dif}{\dif t}\vec{u} + \rho\nabla_r V_\mathrm{g}+\nabla_r P=0,
    \label{eq:momentum_conservation_equation}
\end{equation}
where $\vec{u}$ is the velocity of the fluid, $\rho$ its mass density, $V_\mathrm{g}$ the gravitational potential and $P$ the pressure. Given our approximation of a homogeneous gas temperature the equation of state of the fluid reads
\begin{equation}
    P = c_s^2(T)\rho,
\end{equation}
with $c_s(T)$ being the temperature-dependent speed of sound. \revision{Note that using this approximation renders our treatment of baryonic dynamics equivalent to that used in existing Eulerian analytic descriptions of DM-baryon mixtures, e.g.~\cite{,nusser_analytic_2000,matarrese_growth_2002,shoji_third-order_2009}. This equivalence has also been explicitly demonstrated in \cite{geiss_resummed_2019}. For non-resummed KFT, we have already shown how to go beyond this approximation by including the energy conservation equation, allowing to describe the full ideal gas dynamics with spatial temperature fluctuations \cite{viermann_model_2018}. In future work, we will also explore this in RKFT.}

While the gradients in \eqref{eq:momentum_conservation_equation} are taken with respect to the physical coordinate $\vec{r}$, it is more convenient to work with comoving coordinates $\vec{q} = \vec{r}/a$, with $a$ being the cosmological scale factor normalised to unity today, $a_0=1$. Additionally, it proves useful, to use $\eta \coloneqq \ln{a/a_\im}$ as the time coordinate, where $a_\mathrm{i}$ is the initial scale factor and accordingly $\eta_\im = 0$. Following the derivation in \cite{geiss_resummed_2019} with minor modifications to account for the different normalisation of the scale factor,\footnote{The only differences to the expressions derived in \cite{geiss_resummed_2019}, where we used the normalisation $a_\im = 1$ instead, are the appearance of today's value of the Hubble function, $H_0$, rather than its initial value in \eqref{scalefunction} and \eqref{c_p} as well as the additional $g(0)$ factors in \eqref{eq:cosmo_position_momentum_propagator} and \eqref{scalefunction} to \eqref{b potential}.} we find that the components of the retarded Green's function \eqref{MPH: propagator} read
\begin{align}
    g_{qq}(\eta, \eta') &= \uptheta(\eta - \eta') \,, \\
    g_{qp}(\eta, \eta') &= \uptheta(\eta - \eta') \, \int_{\eta'}^\eta \dif \bar{\eta} \, \frac{g(0)}{g(\bar{\eta})} \,,
    \label{eq:cosmo_position_momentum_propagator} \\
    g_{pq}(\eta, \eta') &= 0 \,, \\
    g_{pp}(\eta, \eta') &= \uptheta(\eta - \eta') \,,
\end{align}
which defines the free motion of both dark and baryonic matter particles. Here, we have introduced the scale function
\begin{equation}
    g(\eta)\coloneqq a(\eta)^2 \, \frac{H(\eta)}{H_0} \,,
    \label{scalefunction}
\end{equation}
depending on the ratio of the Hubble function $H$ and its \revision{value today $H_0$}. Note that the canonically conjugate particle momenta are then given by \begin{equation}
    \vec{p} = \frac{g(\eta)}{g(0)} \, \frac{\dif \vec{q}}{\dif \eta} \,.
\end{equation}

Following \cite{geiss_resummed_2019}, the interaction potentials of the individual particles, appearing in \eqref{eq:total_interaction_potential}, read
\begin{align}
	\revision{v^{\alpha\dm}}(\vec{k},\eta) &= - \frac{a(\eta)}{g(\eta) \, g(0)} \, \frac{C^\alpha_{\mathrm{g}} }{k^2} \label{dm potential},\\
	\revision{v^{\alpha\ba}}(\vec{k},\eta) &= \frac{a(\eta)}{g(\eta) \, g(0)} \, \bigg[-\frac{C^\alpha_{\mathrm{g}} }{k^2}\exp\bigg\{-\frac{3}{4}\sigma_0^2k^2\bigg\} + \delta^{\alpha\ba} \, C_{\mathrm{p}}(\eta) \, a(\eta) \, \exp\bigg\{-\sigma_0^2k^2\bigg\} \bigg]  \label{b potential}
\end{align}
in Fourier space, where we defined the parameters
\begin{align}
C^\alpha_\mathrm{g} &\coloneqq \frac{3 \, \Omega_{\mathrm{m},0}^\alpha}{2 \, \bar{n}^\alpha} \,, \label{c_g} \\
C_\mathrm{p}(\eta) &\coloneqq \frac{c_s^2\bigl(T(\eta)\bigr)}{H_0^2 \, \bar{n}^\ba} \,, \label{c_p}
\end{align}
with $\bar{n}^\alpha$ denoting the mean comoving number density and $\Omega_{\mathrm{m},0}^\alpha$ today's dimensionless matter density parameter of $\alpha$ particles, respectively. The potential \eqref{dm potential} for the microscopic DM particles directly follows from the Poisson equation for the gravitational potential. For the mesoscopic baryonic particles, the fluid equations have been projected onto the individual mesoscopic particle contributions, resulting in equations of motion analogous to those of the DM particles with a modified interaction potential which includes an additional pressure term \cite{viermann_model_2018, geiss_resummed_2019}. The parameter $\sigma_0$ denotes the comoving mesoscopic scale for which we will use the limit of ideal hydrodynamics $\sigma_0\rightarrow 0$ from here on, which is a valid approximation as we are interested in scales much larger than the mean free path.

In addition, we can use the fact that in the thermodynamic limit the values of the mean particle number densities do not have any physical meaning on their own. Only the mean mass densities are measurable. We are thus free to set the masses of baryonic and dark matter particles equal, yielding
\begin{equation}
    \bar{n}_\alpha = \frac{\Omega_{\mathrm{m},0}^\alpha}{\Omega_{\mathrm{m},0}} \, \bar{n} \,, \label{mean_density_relation}
\end{equation}
where $\Omega_{\mathrm{m},0} = \Omega_{\mathrm{m},0}^\ba + \Omega_{\mathrm{m},0}^\dm$ and $\bar{n} = \bar{n}^\ba + \bar{n}^\dm$. Using this relation, the potential coefficients \eqref{c_g} and \eqref{c_p} become
\begin{align}
    C_\mathrm{g}^\alpha &= \frac{3 \, \Omega_{\mathrm{m},0}}{2 \, \bar{n}}. \label{c_g with total density} \\
    C_\mathrm{p}(\eta) &= \frac{\Omega_{\mathrm{m},0} \, c_s^2\bigl(T(\eta)\bigr)}{\Omega_{\mathrm{m},0}^\ba \, \bar{n} \, H_\mathrm{0}^2}. \label{c_p with total density}
\end{align}
Note that the equal mass of both particle species automatically implies equal single-particle gravitational potentials. $C_\mathrm{g}^\ba = C_\mathrm{g}^\dm$.

\subsection{Evolution of gas pressure}
Let us now take a closer look at the pressure potential coefficient $C_\mathrm{p}$ in \eqref{c_p with total density}. To calculate the sound velocity we assume that the baryonic matter in the Universe consists of monoatomic ideal hydrogen gas which indeed makes up most of the baryonic matter. For \revision{an ideal gas} the sound velocity is given by
\begin{equation}
    c_s\bigl(T(\eta)\bigr) = \sqrt{\frac{\gamma \, k_\mathrm{B} \, T(\eta)}{m}},
    \label{eq:isothermal_sound_velocity}
\end{equation}
with $\gamma$ being the adiabatic index of the gas, $k_\mathrm{B}$ the Boltzmann constant and $m$ the mass of the particles that make up the gas. In a monoatomic gas, the particles have $f = 3$ degrees of freedom, fixing the adiabatic index to
\begin{equation}
    \gamma = \frac{f+2}{f} = \frac{5}{3}.
\end{equation}
The mass of a hydrogen atom can be safely approximated by the mass of a proton $m_\mathrm{P}=1.67\cdot 10^{-27}$kg since the electron mass is negligible. Altogether, $C_\mathrm{p}$ thus takes the form
\begin{equation}
    C_\mathrm{p}(\eta) = \frac{5 \, \Omega_{\mathrm{m},0} \, k_\mathrm{B} \, T(\eta)}{3 \, \Omega_{\mathrm{m},0}^\ba \, \bar{n} \, m_\mathrm{P} H_\mathrm{0}^2}. \label{c_p specialised}
\end{equation}
\revision{Note that baryonic matter in our model is treated independently from photons. However, even after recombination this is not strictly the case at high redshifts when collisions between photons and baryons are still frequent. This in turn leads to a drag experienced by the baryons and changes the actual speed of sound \cite{ma_cosmological_1995,blas_cosmic_2011}. Describing this interaction accurately requires to treat the photons as a separate particle species, but currently it is not possible to describe relativistic particles within KFT. To capture the approximate effect of photons on the speed of sound, we will account for their influence on the mean baryonic gas temperature.}

The scale below which the effects of the baryonic pressure play a role is characterised by the baryonic Jeans length $\lambda_\mathrm{J}$. The associated Jeans wavenumber $k_\mathrm{J} = 2 \pi / \lambda_\mathrm{J}$ is obtained by finding the wavenumber for which the baryon interaction potential \eqref{b potential} vanishes,
\begin{equation}
    k_\mathrm{J}(\eta) = \sqrt{\frac{C_\mathrm{g}^\ba}{C_\mathrm{p}^\ba(\eta) \, a(\eta)}} = \sqrt{\frac{9 \, \Omega_{\mathrm{m},0}^\ba \, H_0^2 \, m_\mathrm{P}}{10 \, a(\eta) \, k_\mathrm{B} \, T(\eta)}} \,.
    \label{eq:jeans_wavenumber}
\end{equation}
Its time dependence is determined by the evolution of the baryonic gas temperature $T(\eta )$ after recombination, which is illustrated in Figure 15 of \cite{zaroubi_epoch_2013}. It can be divided into the following three regimes:
\begin{enumerate}
\item $1100 \gtrsim z \gtrsim 200$: Photons scatter off the baryonic matter often enough for it to follow the CMB temperature. Therefore, the temperature scales like $a^{-1}$.
\item $200 \gtrsim z \gtrsim 30$: In this regime, the baryon temperature evolves independently from the photons. Since we have adiabatic expansion of the Universe and we are dealing with non-relativistic matter the temperature scales like $a^{-2}$.
\item $30 \gtrsim z $: At this point reionization begins and the baryons heat up again. In addition, baryons will also be heated by the gravitational collapse of structures. However, for the qualitative analysis in this paper we only consider the linear evolution of structures and also neglect the effect of reionization. Hence, we assume the baryons to keep cooling adiabatically during this epoch.
\end{enumerate}
Overall, we thus assume the following evolution of the baryonic gas temperature,
\begin{equation}
    T(\eta) = \begin{cases}
        \frac{T_\mathrm{CMB}}{a(\eta)} &\text{if $z \geq 200$,} \\
        \frac{T_\mathrm{CMB}}{201 \, a(\eta)^2} &\text{else,}
    \end{cases}
    \label{temperature evolution}
\end{equation}
where $T_\mathrm{CMB} = 2.725 \, \mathrm{K}$ is the CMB temperature today. This implies that $k_\mathrm{J}$ stays constant up to $z=200$ and grows with the square root of the scale factor from then on.

\section{Linearly evolved power spectra \label{results}}
With the dynamics of both dark and baryonic matter fixed, we can now proceed to analyse the linear evolution of the baryonic, dark and total matter spectra, computed from the tree-level RKFT propagator. \revision{The details of this computation are explained in \autoref{App:linear_power_spectra_computation}.} For this, we consider a flat $\Lambda$CDM cosmology and set the current dimensionless density parameters to $\Omega_{\mathrm{m},0} = 0.3$ and $\Omega_{\Lambda,0} = 0.7$. In addition, we assume baryons to account for 16\,\% of the total matter, $\Omega_{\mathrm{m},0}^\ba / \Omega_{\mathrm{m},0} = 0.16$, and fix the Hubble constant to $H_0 = 70 \, \mathrm{km/s} \, \mathrm{Mpc}^{-1}$. All following calculations use an initial redshift of $z_\im = 1000$, corresponding to a time shortly after recombination.

\subsection{Qualitative analysis}
\begin{figure}
  \centering
  \begin{minipage}{0.49\textwidth}
    \includegraphics[width=\textwidth]{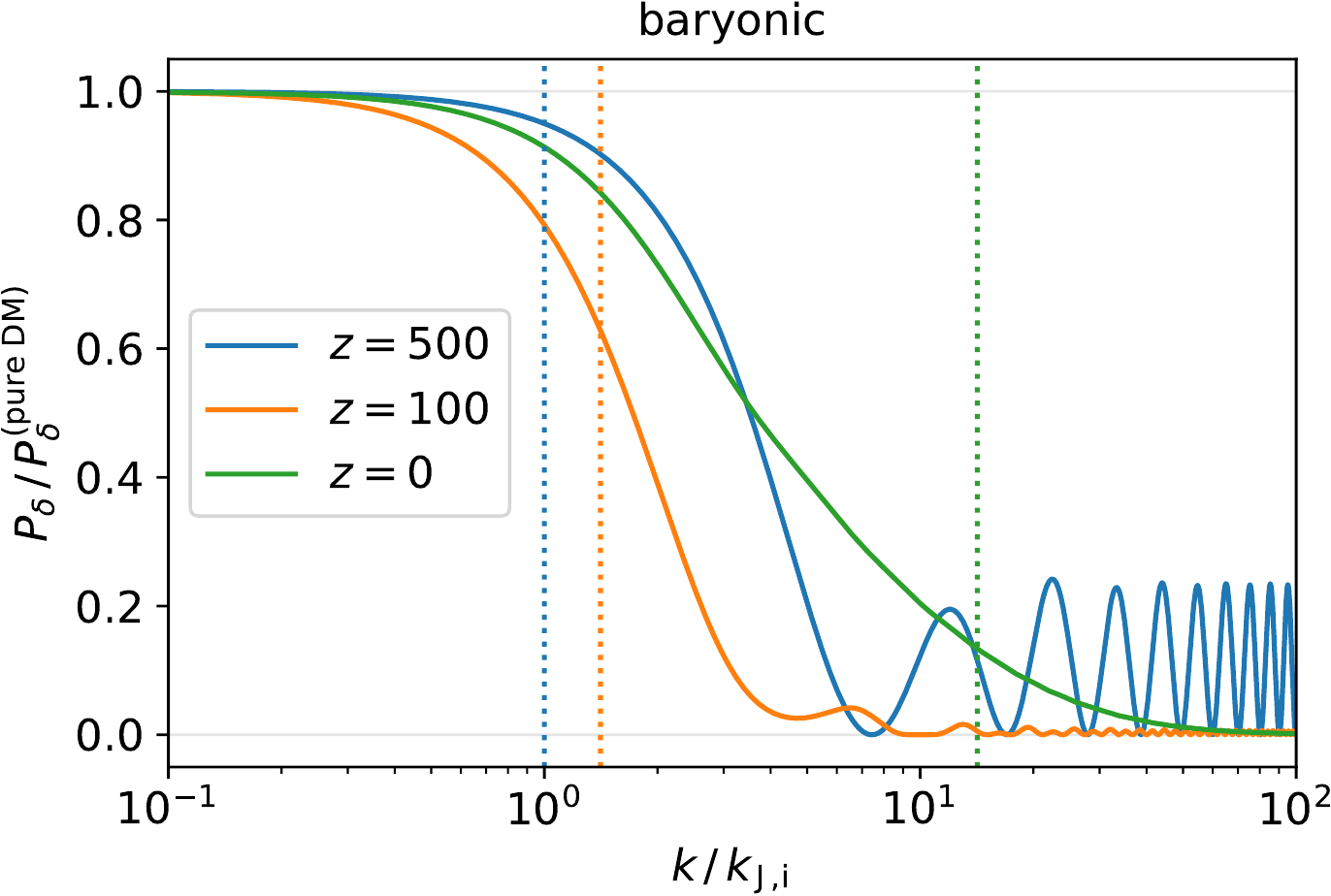}
  \end{minipage}
  \hfill
  \begin{minipage}{0.49\textwidth}
    \includegraphics[width=\textwidth]{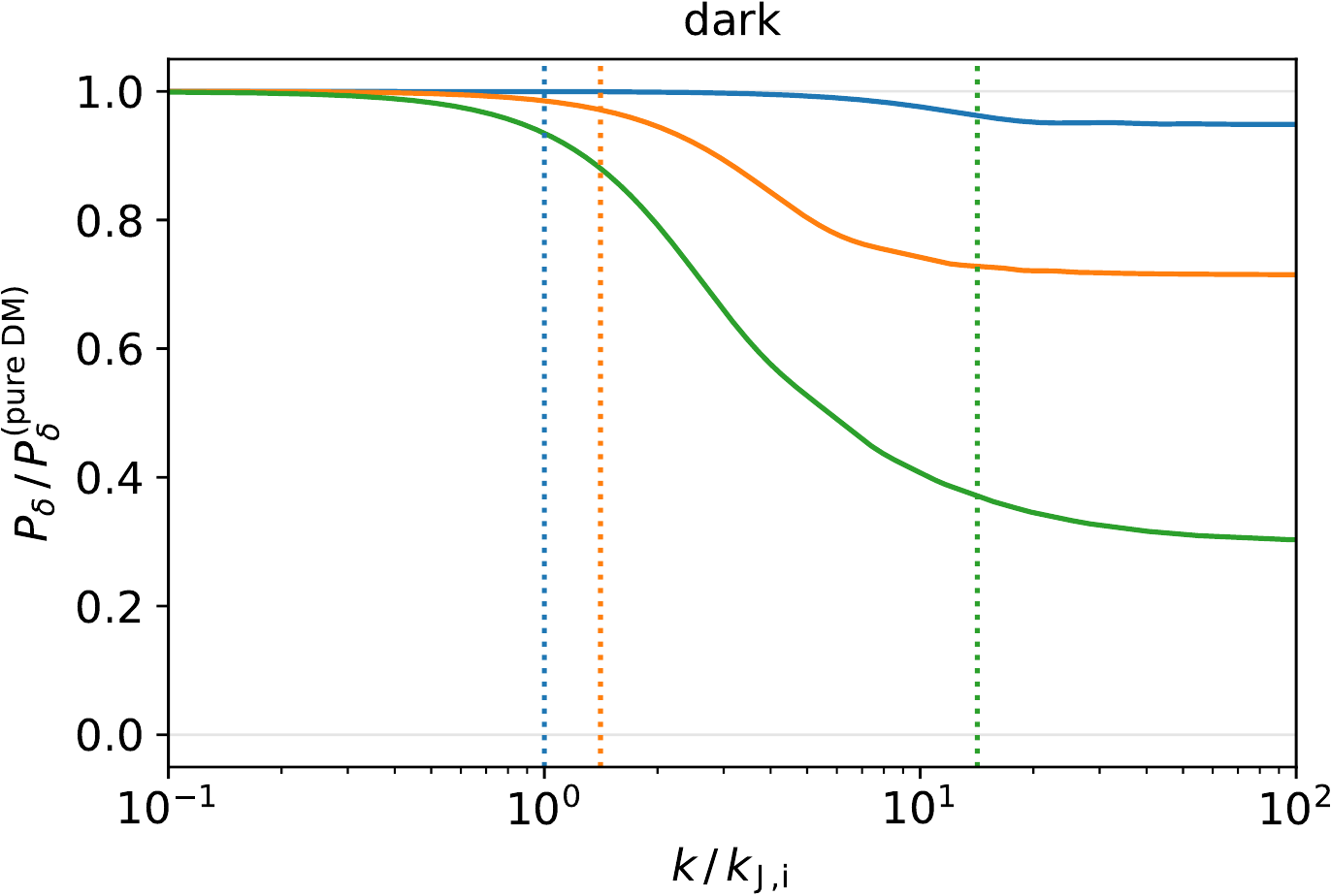}
  \end{minipage}
  \\[0.75\baselineskip]
  \begin{minipage}{0.49\textwidth}
    \includegraphics[width=\textwidth]{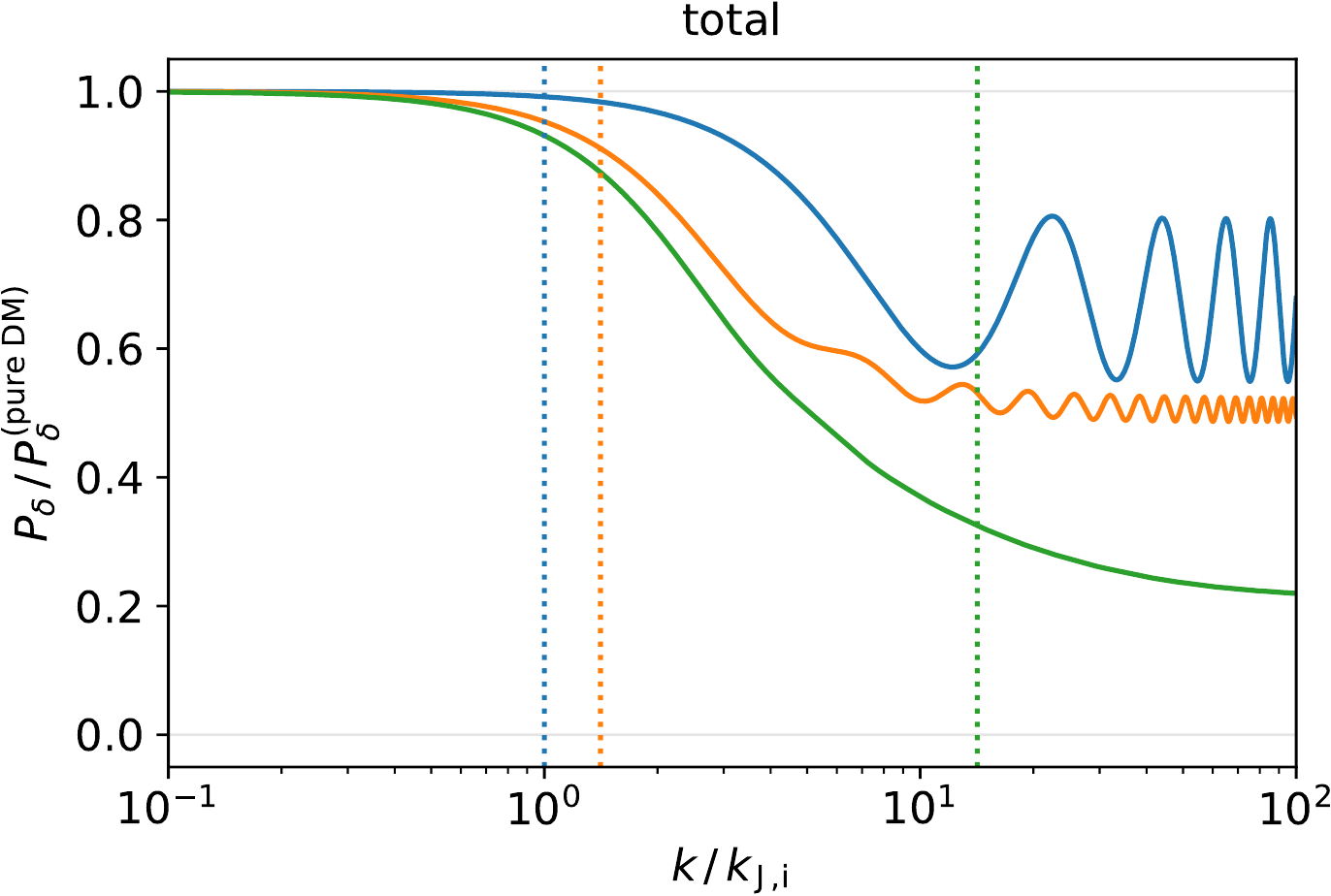}
  \end{minipage}
  \caption{Linearly evolved tree-level RKFT power spectra of baryonic (top left), dark (top right) and total matter (bottom) divided by the spectrum in a pure DM scenario, evaluated at the redshifts 500 (blue), 100 (orange) and 0 (green). The horizontal axis shows the wavenumber in units of the intial baryonic Jeans wavenumber $k_\mathrm{J,i}$ and the current Jeans wavenumbers at the different redshifts are marked by vertical dotted lines in the same colors as the spectra. On wavenumbers larger than $k_\mathrm{J,i}$, the gas pressure strongly suppresses the baryonic spectrum and creates acoustic oscillations. Because of the gravitational coupling between baryons and DM these features are also present in a less pronounced way in the dark and total matter spectra.
  \label{fig:TimeEvolutionFull}}
\end{figure}

Let us first consider the case that baryons and DM follow exactly the same initial power spectrum. While this is actually not the case, it allows us to focus only on the effect that the baryonic gas dynamics has on the linear structure formation. For this, we compute the ratios of the baryonic, dark and total matter spectra, obtained from the RKFT tree-level result \eqref{nn-component of the propagator}, to the linear spectrum of a pure DM system. Note that these ratios are independent of the form of the initial spectrum. In \autoref{fig:TimeEvolutionFull} we plot the results for three exemplary redshifts, $z = 500$, 100 and 0, over a range of scales around the initial Jeans wavenumber $k_\mathrm{J,i} = k_\mathrm{J}(0)$. We additionally mark the current Jeans wavenumbers at the three redshifts by vertical dotted lines.

As expected, on wavenumbers smaller than $k_\mathrm{J,i}$ gravity dominates and thus the power spectra of any matter species evolve at the same rate as in the pure DM scenario. At larger wavenumbers, on the other hand, the baryon-baryon spectrum quickly drops significantly below the pure DM spectrum since the repulsive effect of the gas pressure suppresses the growth of baryonic structures. In addition, the small-scale baryonic spectrum develops oscillations caused by the fact that the counteracting gravitational and pressure forces create wave-like perturbations. Over time the amplitude of these oscillations decreases relative to the linearly growing pure DM spectrum. At late times the current Jeans wavenumber $k_\mathrm{J}(\eta)$ starts to increase once the baryons fully decouple from the photons and cool adiabatically, which happens at $z=200$ in our model. As soon as a Fourier mode falls below $k_\mathrm{J}(\eta)$ it effectively evolves purely gravitationally again. The baryonic spectrum then grows relative to the pure DM spectrum because baryons do not only experience the gravitational attraction from other baryons but also that from the dominant DM species.

Since the resummed KFT propagator includes the full mutual gravitational interactions between baryonic and dark matter, the influence of the baryonic pressure is also seen in the DM spectrum. The dominant effect of this is a suppression in DM structure growth above $k_\mathrm{J,i}$, though not as pronounced as in the baryonic spectrum. As a secondary effect, we can also observe very small oscillations in the small-scale DM spectrum. These phenomena are mainly caused by the lack of baryonic structures on these scales which in turn decreases the overall gravitational attraction experienced by DM. Additionally, the baryons are dragging a fraction of the DM with them through gravitational interactions as they themselves are pushed apart due to pressure forces. The total matter spectrum expectedly shows a behaviour that is a mix between the spectra of the two particles species, dominated by the larger DM contribution. All in all, we thus find that the RKFT tree-level result for the power spectra displays exactly the behaviour one would expect from the linear evolution of a mixed system of baryonic and dark matter.

\subsection{Comparison to Boltzmann solver}
Now that we have qualitatively investigated the influence of the baryonic pressure on structure growth, we want to verify the validity of our description quantitatively. This requires to account for appropriate initial conditions for both baryonic and dark matter. In contrast to DM, baryons are interacting with photons during the epoch of radiation domination and the early phases of matter domination. The largest-scale traces of these non-gravitational interactions are the baryon acoustic oscillations (BAOs), whose first peak in the power spectrum appears between $k = 0.01$ and $0.1 \,h\,\mathrm{Mpc}^{-1}$. Beyond that, the baryonic spectrum is subject to Silk damping, an exponential suppression of baryonic structures at smaller scales caused by the increasing diffusion of photons during the epoch of recombination \cite{silk_cosmic_1968}.

Apart from the effect on the mean gas temperature, we are not explicitly accounting for the coupling between baryons and photons in this paper, though, but only for the gas pressure itself. Therefore, we cannot expect an accurate description of the BAO evolution and will only focus on wavenumbers $k \geq 1 \, h/\mathrm{Mpc}$ for which BAOs do not play a role and the initial baryonic spectrum is just strongly suppressed relative to the DM spectrum by Silk damping. The formation and evolution of BAOs within RKFT, using an effective model for the baryon-photon interactions, will be investigated in a follow-up paper. Note that for our purposes here, restricting the analysis to these high wavenumbers presents no problem since the effects of the baryonic gas pressure on the linear evolution only come into play for $k \gtrsim k_\mathrm{J,i} \approx 140 \, h/\mathrm{Mpc}$. This changes, however, once nonlinear corrections are considered. Then baryons will be heated significantly by the collapse of structures, bringing $k_\mathrm{J}$ down below $1 \, h/\mathrm{Mpc}$ or even $0.1 \, h/\mathrm{Mpc}$ in dense clusters. But this goes beyond the scope of the current paper.

\begin{figure}
  \centering
  \begin{minipage}{0.49\textwidth}
    \includegraphics[width=\textwidth]{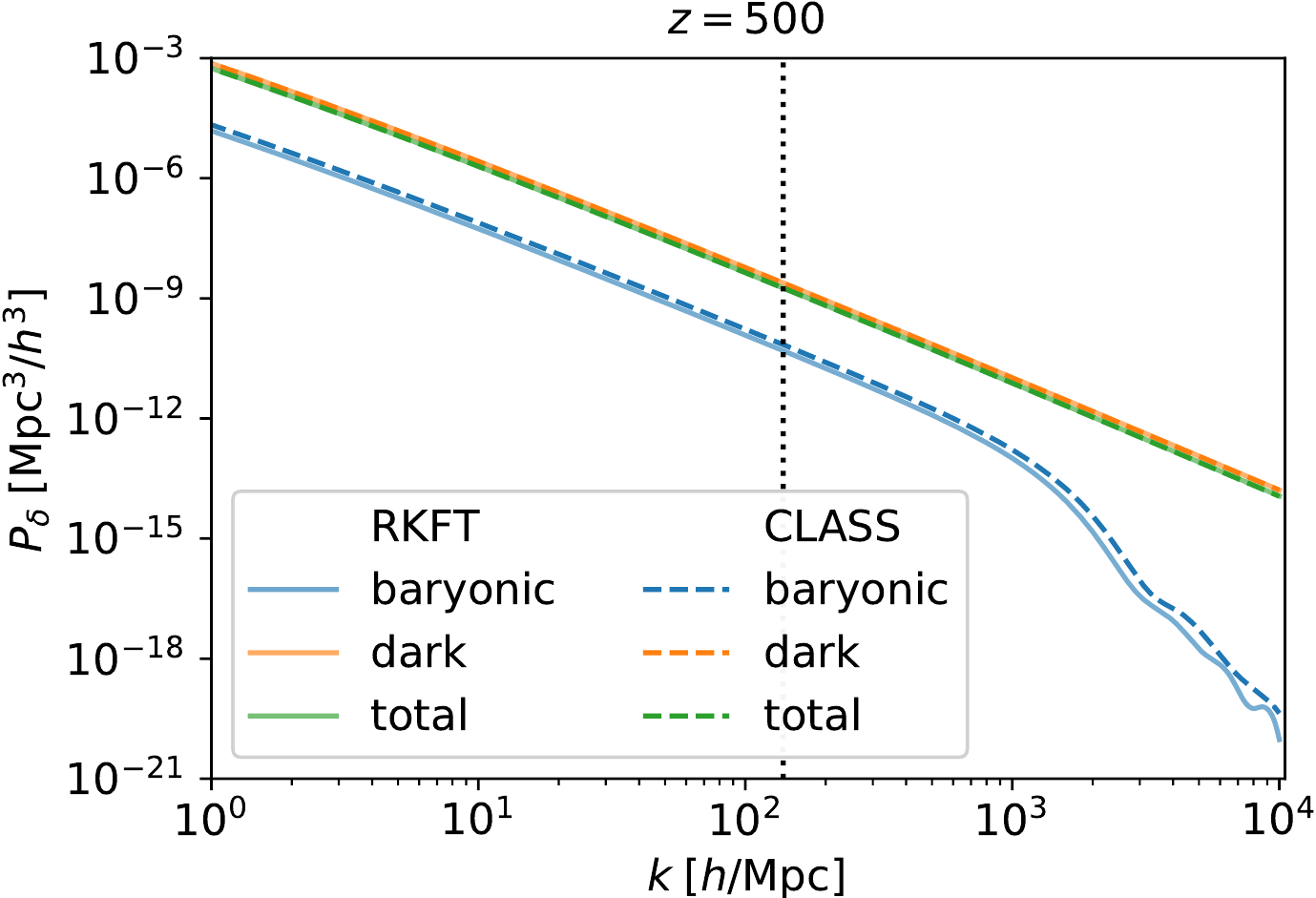}
  \end{minipage}
  \hfill
  \begin{minipage}{0.49\textwidth}
    \includegraphics[width=\textwidth]{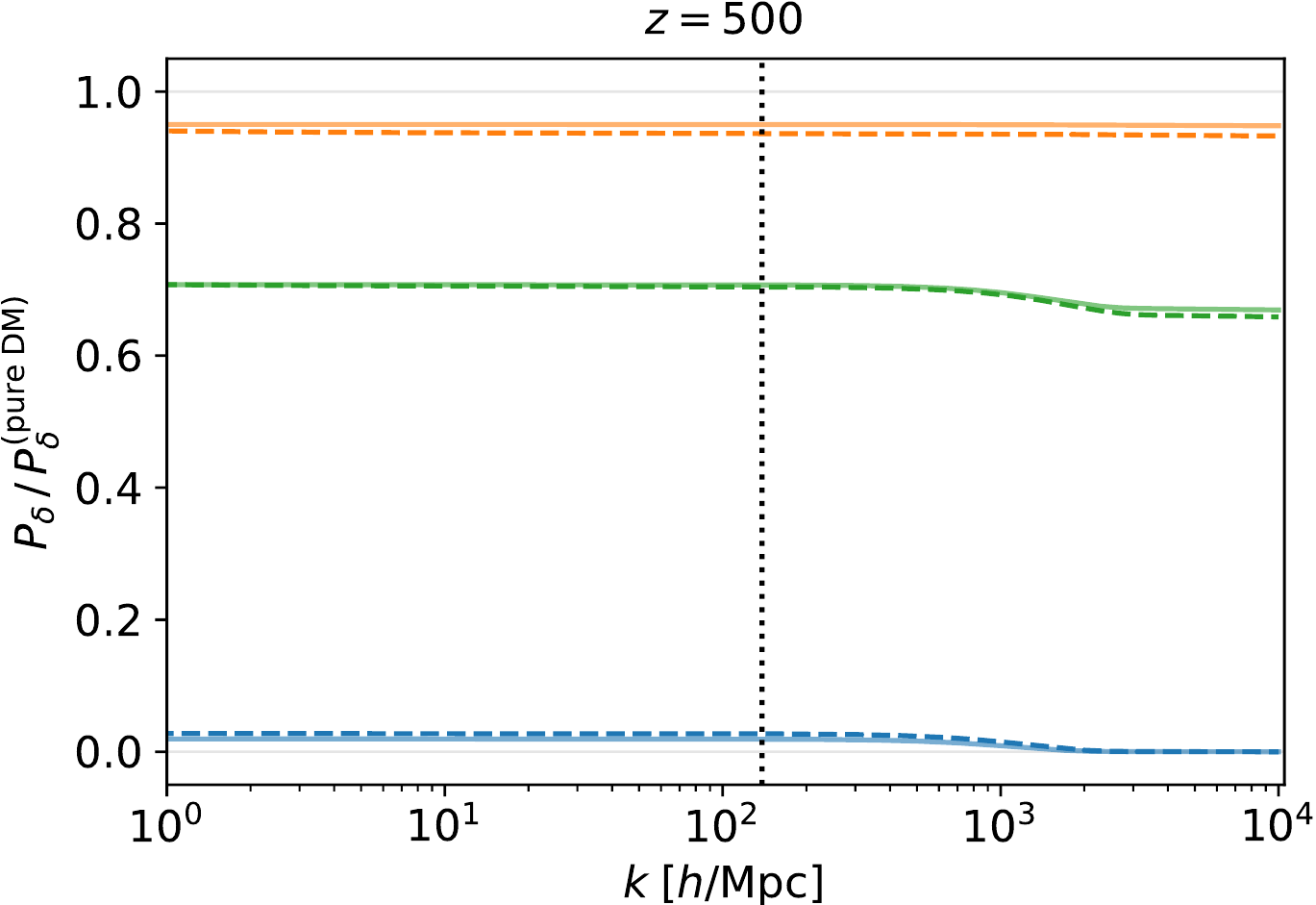}
  \end{minipage}
  \\[0.75\baselineskip]
  \begin{minipage}{0.49\textwidth}
    \includegraphics[width=\textwidth]{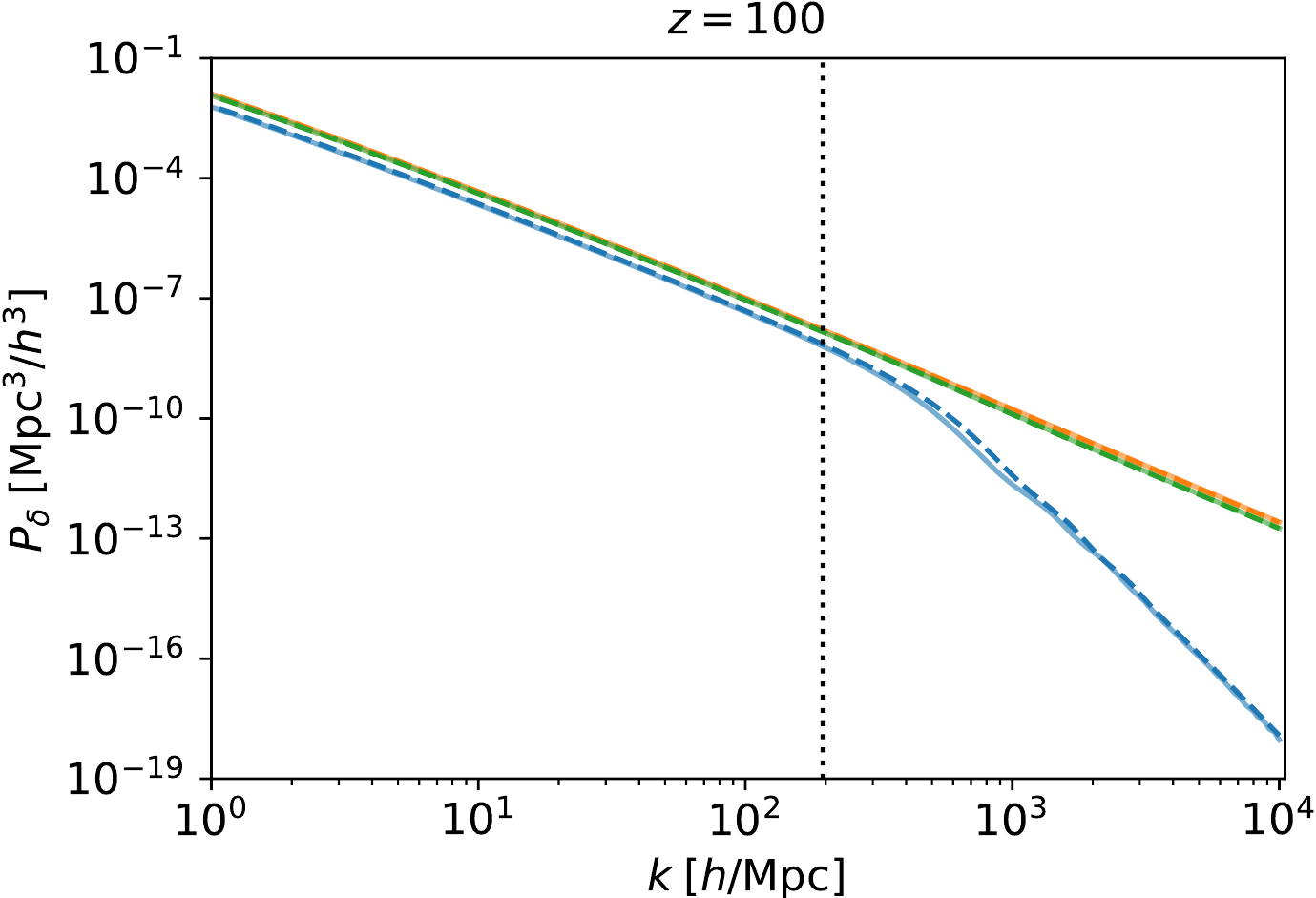}
  \end{minipage}
  \hfill
  \begin{minipage}{0.49\textwidth}
    \includegraphics[width=\textwidth]{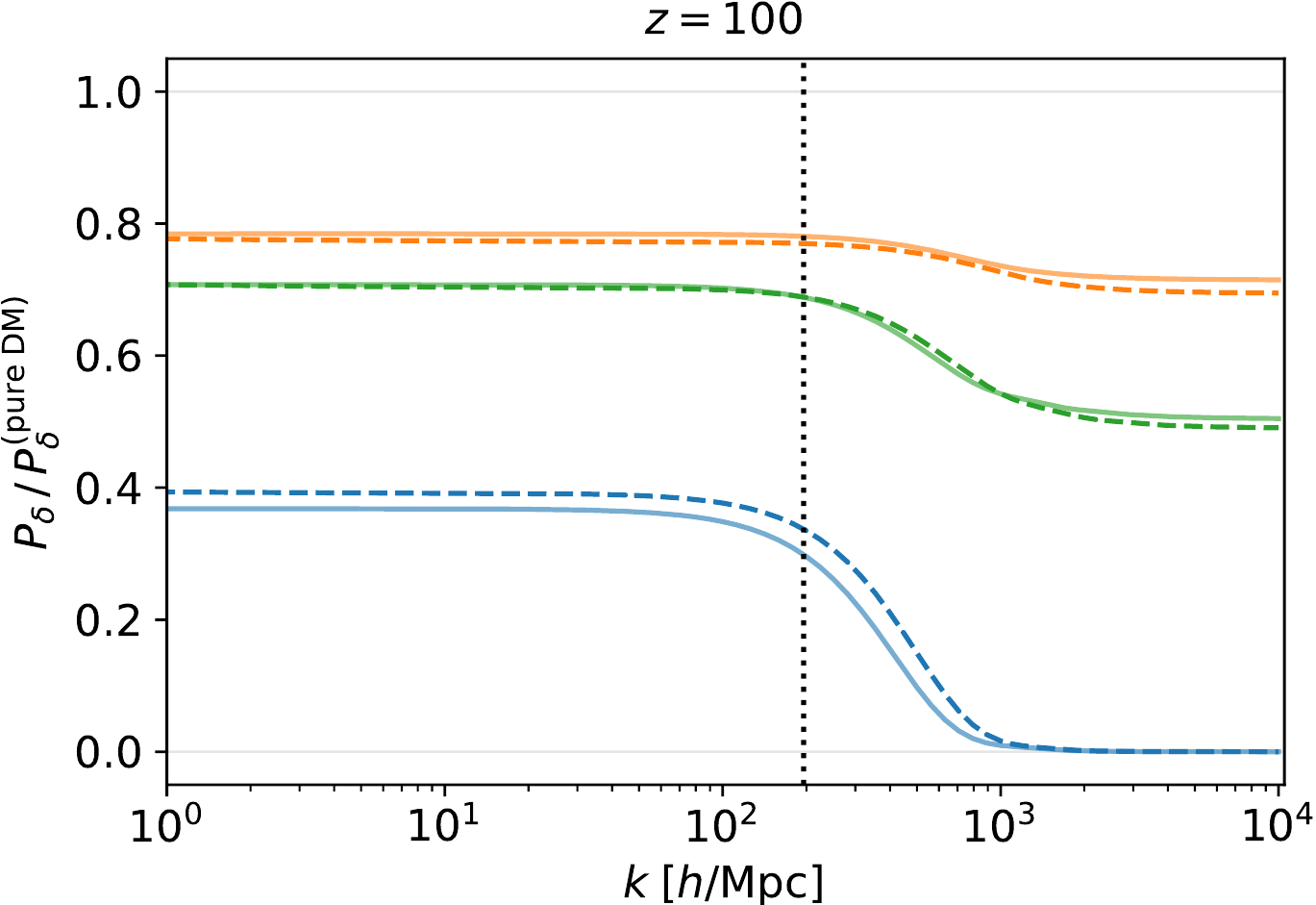}
  \end{minipage}
  \\[0.75\baselineskip]
  \begin{minipage}{0.49\textwidth}
    \includegraphics[width=\textwidth]{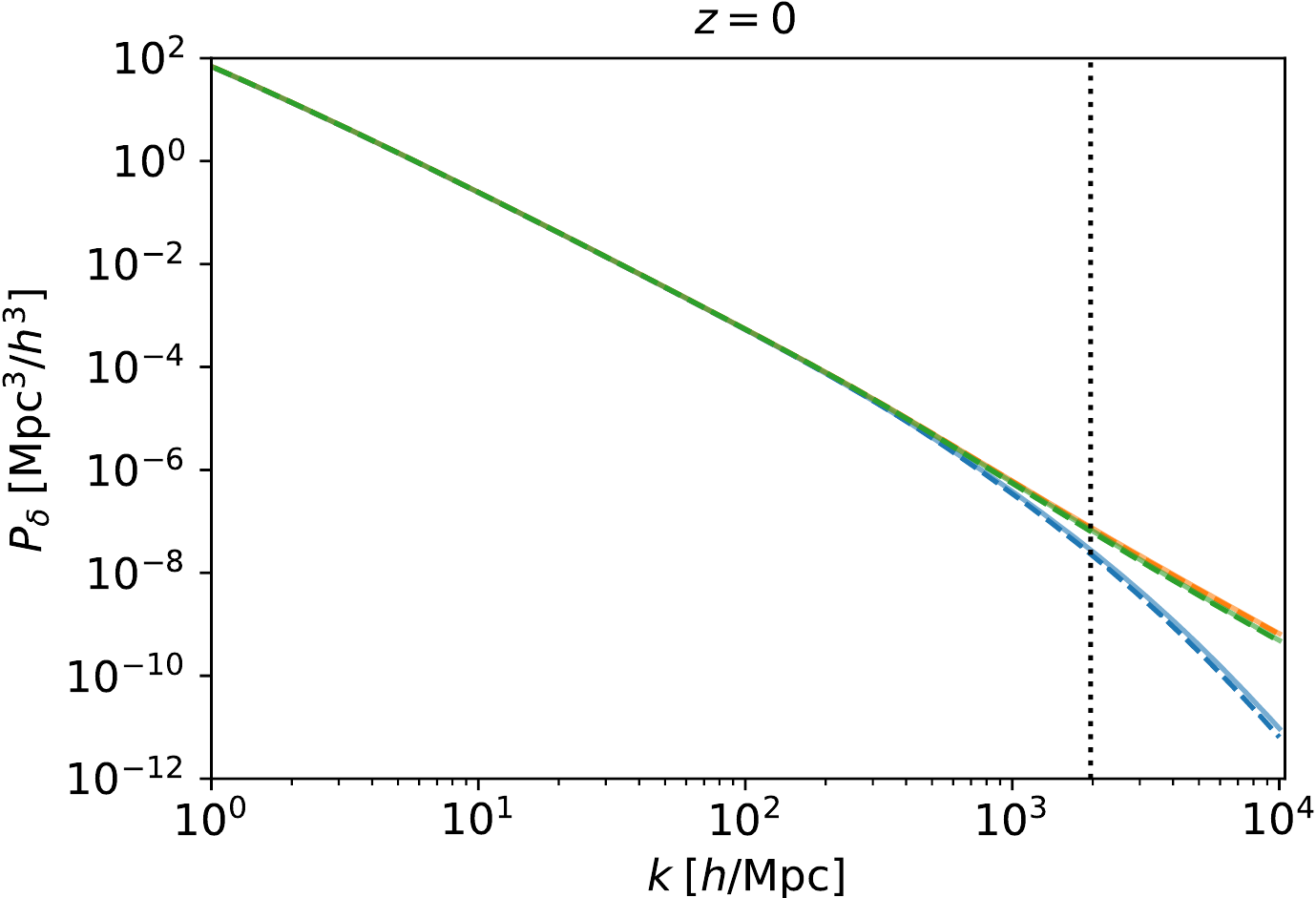}
  \end{minipage}
  \hfill
  \begin{minipage}{0.49\textwidth}
    \includegraphics[width=\textwidth]{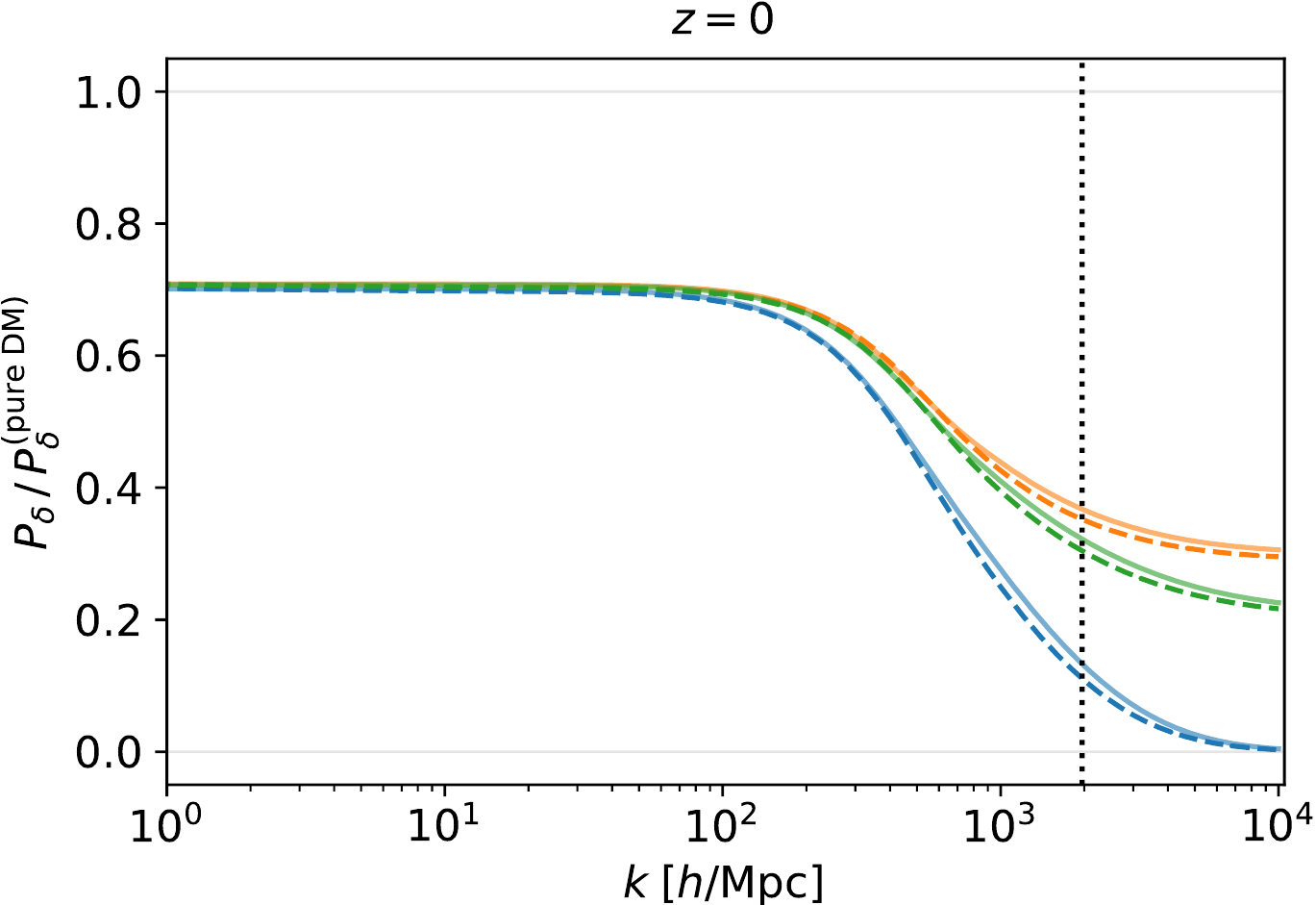}
  \end{minipage}
  \caption{Comparison between the linearly evolved power spectra of baryonic (blue), dark (orange) and total matter (green) obtained from the tree-level result of our analytical RKFT framework (solid) and the Boltzmann solver CLASS \cite{blas_cosmic_2011} (dashed). From top to bottom the rows show the results for redshifts 500, 100 and 0. On the left we plot the spectra themselves, while on the right they are divided by the spectrum in a pure DM scenario. In each panel, the vertical dotted black line marks the current baryonic Jeans wavenumber $k_\mathrm{J}$. For all redshifts and types of matter we find good agreement between the RKFT and CLASS spectra, with the slight deviations being expected to originate from the approximations made in our calculations. \label{fig:comparison_CLASS}}
\end{figure}

As a reference to compare our approach against we use the numerical Boltzmann solver CLASS \cite{blas_cosmic_2011} to compute the power spectra of baryons and DM at the redshifts $z = 1000$, 500, 100 and 0, assuming a spectral index of unity and a normalisation of $\sigma_8 = 0.8$ today. We then use the \revision{auto- and cross-spectra $P_\delta^{\alpha \gamma}$} at $z_\im = 1000$ as initial conditions and compute the linearly evolved spectra at the later redshifts within tree-level RKFT. The results are compared to the CLASS spectra in \autoref{fig:comparison_CLASS}. In the left panel we plot the spectra themselves, in the right panel we divide them by the linear spectrum for pure DM again.

For all redshifts, we find good agreement between our results and the CLASS spectra, both with regards to the amplitudes of the spectra as well as the wavenumbers beyond which the suppression due to baryonic pressure sets in. While there are small deviations, these seem well within reason, considering the approximations we \revision{made. Assuming a spatially constant gas temperature with a simplified model for its time evolution introduced small inaccuracies in the baryonic speed of sound and thus the Jeans wavenumber $k_\mathrm{J}$. Accordingly, this approximation is likely to cause the slight deviations in the wavenumber above which pressure starts to suppress structure growth. Neglecting the drag due to scattering photons after $z_\im$, on the other hand, should explain the small mismatch in the amplitudes even on wavenumbers smaller than $k_\mathrm{J}$.}

\FloatBarrier
\section{Conclusion and outlook \label{conclusion}}
Building on earlier works on Kinetic Field Theory (KFT) \cite{bartelmann_microscopic_2016, bartelmann_kinetic_2017, bartelmann_analytic_2017}, its resummation scheme (RKFT) \cite{lilow_resummed_2019} and a description of baryonic matter through effective particles (MPH) \cite{viermann_model_2018, geiss_resummed_2019}, we have expanded the KFT formalism to systems of two different particle species and used it to calculate the linear evolution of a mixture of baryonic and dark matter in a cosmological framework. For doing so, we first formulated an appropriate generating functional incorporating the dynamics of two coupled particle species. By generalizing the proceedings of \cite{fabis_kinetic_2018}, we calculated the free two point cumulants. Using these results, we showed how the density power spectra for the two types of matter can straightforwardly be computed within the RKFT framework. To test the validity of our approach, we then computed these spectra in lowest perturbative order, corresponding to linear evolution, within a $\Lambda$CDM cosmology, with the initial redshift set shortly after the epoch of recombination. In our model, both baryons and DM interact gravitationally, while only the baryons experience an additional gas pressure. We further use the approximation of a spatially constant but time-dependent gas \revision{temperature. At early times, when photons still frequently scatter off the baryons, the baryons follow the photon temperature, before they eventually cool adiabatically. Apart from this effect, we do not account for any coupling between baryons and photons.}

We find that our description of the coupled baryon-DM system captures all the effects expected to arise in linear evolution. The gas pressure leads to a suppression of structure growth and the formation of acoustic oscillations in the baryonic spectrum on scales smaller than the Jeans length. Due to the gravitational interactions between both types of matter, these effects are also present in a weakened form in the spectrum of the pressureless DM as well as the total matter spectrum. For a quantitative analysis, we compared our results to the spectra computed with the numerical Boltzmann solver CLASS \cite{blas_cosmic_2011}. Thereby, we focused on wavenumbers above $1 \, h/\mathrm{Mpc}$, where Silk damping implies an essentially vanishing initial baryonic spectrum. Within the limits of our approximations, we find good quantitative agreement, which also implies that the formation \revision{and subsequent growth} of initially absent baryonic structures, due to the gravitational coupling to DM, is correctly captured by the resummed microscopic interactions in RKFT.

We consider this a clear confirmation of the feasibility and validity of our approach to describe structure formation in a coupled system of dark and baryonic matter by a combination of RKFT and MPH. Of course, an accurate description of the linear evolution of this mixed system is also possible in approaches building on fluid dynamics, as demonstrated for example in \cite{nusser_analytic_2000,matarrese_growth_2002,shoji_third-order_2009}. However, once we proceed to include nonlinear contributions, DM particles can no longer be accurately treated as a fluid since they form crossing streams. With the consistent inclusion of baryonic gas dynamics into the particle-based framework of (R)KFT, we are thus laying the stepping stone for an accurate treatment of structure formation in the full matter content of our Universe.

To proceed towards this goal, future work should focus on the following three aspects:
(i) \revision{Nonlinear} corrections to the resummed power spectrum need to be included. \revision{The first step in this direction will be the computation of 1-loop contributions within the macroscopic perturbation theory. Formally, this proceeds analogously to computing 1-loop corrections in Eulerian Standard Perturbation Theory, but accounting for the different form of the propagators, \eqref{nn-component of the propagator} and \eqref{integral equation for retarded propagator}, and vertices, \eqref{macroscopic_vertices}. Details on the systematic expansion of RKFT in loop orders are presented in \cite{lilow_resummed_2019}.}
(ii) The nonlinear collapse of structures will lead to a heating of baryons that is not well approximated by a spatially constant gas temperature. Extensions to our current treatment thus need to be investigated. \revision{A simple heuristic approach would be to make the gas temperature entering the computation of the power spectrum in \eqref{free_nBv_cumulant_cosmo} wavenumber-dependent in a way that represents the average temperature of structures on the associated length scale. A more rigorous extension consists in explicitly accounting for the energy conservation equation in the baryonic dynamics, thus describing the full ideal gas dynamics. The latter has already been explored in non-resummed KFT in \cite{viermann_model_2018}. It requires to give the mesoscopic particles an enthalpy (or equivalently an internal energy) as an additional intrinsic property, which brings along additional components to the Green's function \eqref{MPH: propagator} as well as additional contributions to the interaction part of the action \eqref{Interacting part of microscopic action}. The next step will be to resum this enthalpy-extended KFT by reformulating it in terms of macroscopic fields, in close analogy to \cite{lilow_resummed_2019}, and to generalise it to the two particle-species case, in the same fashion as described in the present paper.}
(iii) So far, only repulsive pressure effects are taken into account, leading to a suppression of structure growth. By modifying the effective \revision{mesoscopic particle interactions} appropriately, other baryonic effects such as radiative cooling, which enhances structure formation, could be considered. \revision{These modifications could be derived by introducing radiative loss and gain terms into the baryonic energy conservation equation.}

Further developments in these directions will provide a powerful analytical tool to gain deeper physical understanding of the impact baryons have on cosmic structure growth.

\acknowledgments{
    We are grateful for many helpful comments and discussions to Felix Fabis and Celia Viermann. This work was supported  by the Deutsche Forschungsgemeinschaft (DFG, German Research Foundation) under Germany's Excellence Strategy EXC 2181/1 - 390900948 (the Heidelberg STRUCTURES Excellence Cluster) as well as the Heidelberg Graduate School of Physics. DG acknowledges funding by International Max Planck Research Schools (IMPRS). RL acknowledges support by a Technion fellowship.
}

\appendix
\section{Derivation of the initial distribution} \label{App: Polynomial}
In \autoref{Sec: Init Distr} we explain how to construct the initial distribution via a sampling of the macroscopic fields and just state the final result. In this section, we give a few more details on the derivation of that result, focusing on the key generalizations with respect to the case of single particle species presented in \cite{bartelmann_microscopic_2016} and \cite{fabis_kinetic_2018}.

The initial phase-space distribution of the particles can be formulated in terms of the distribution of the macroscopic field as
\begin{equation}
	P(\vec{\boldsymbol{x}}^{(\mathrm{i})}) = \int \mathrm{d}\vec{\boldsymbol{d}} P(\vec{\boldsymbol{x}}^{(\mathrm{i})}|\vec{\boldsymbol{d}})P(\vec{\boldsymbol{d}}),  \label{Ptotal}
\end{equation}
where $\vec{\boldsymbol{d}}$ denotes the data tensor given by
\begin{equation}
	\vec{\boldsymbol{d}} \coloneqq \begin{pmatrix} \boldsymbol{d}^\mathrm{b} \\ \boldsymbol{d}^\mathrm{d} \end{pmatrix}, \qquad \boldsymbol{d}^\alpha \coloneqq \begin{pmatrix} \delta^{\alpha(\mathrm{i})}_j \\ \vec{P}^{\alpha(\mathrm{i})}_j \end{pmatrix} \otimes \vec{e}^\alpha_j.
\end{equation}
Here $\delta^{\alpha(\mathrm{i})}_j \coloneqq 	\delta^{\alpha(\mathrm{i})}\bigl(\vec{q}_j^{(\mathrm{i})}\bigr)$ and $\vec{P}^{\alpha(\mathrm{i})}_j \coloneqq 	\vec{P}^{\alpha(\mathrm{i})}\bigl(\vec{q}^{(\mathrm{i})}_j\bigr)$ correspond to the macroscopic initial density contrast $\delta^{\alpha\mathrm{(i)}}(\vec{q})$, defined by $n^{\alpha(\mathrm{i})}(\vec{q}) = \bar{n}^\alpha \big(1 + \delta^{\alpha(\mathrm{i})}(\vec{q}) \big)$, and the initial macroscopic momentum field $\vec{P}^{\alpha\mathrm{(i)}}(\vec{q})$ evaluated at the initial position of the particle $j^\alpha$. The conditional phase-space distribution $P(\vec{\boldsymbol{x}}^{(\mathrm{i})}|\vec{\boldsymbol{d}})$ describes a Poisson sampling process from a specific realization of the initial macroscopic fields,
\begin{equation}
	P(\vec{\boldsymbol{x}}^{(\mathrm{i})}|\vec{\boldsymbol{d}}) = \prod_{\alpha\in\{\mathrm{b,d}\}} \, \prod_{j=1}^{N^\alpha} \frac{1}{N^\alpha} \, n^{\alpha(\mathrm{i})}\bigl(\vec{q}_j^{\alpha(\mathrm{i})}\bigr) \; \delta_\mathrm{D} \! \left( \vec{p}^{\alpha(\mathrm{i})}_j - \revision{\vec{P}}^{\alpha(\mathrm{i})}\bigl(\vec{q}_j^{\alpha(\mathrm{i})}\bigr) \right). \label{Pmicro}
\end{equation}
For the macroscopic fields a Gaussian distribution is assumed, i.e.
\begin{equation}
	P(\vec{\boldsymbol{d}}) = \Bigl((2\pi)^{4(N^\mathrm{b}+N^\mathrm{d})}\det \boldsymbol{C}\Bigr)^{-1/2} \exp\left\{ -\frac{1}{2} \, \vec{\boldsymbol{d}}^\intercal \, \boldsymbol{C}^{-1} \, \vec{\boldsymbol{d}} \right\} ,\label{Pmacro}
\end{equation}
where the covariance matrix 
\begin{equation}
	\boldsymbol{C} \coloneqq \left\langle \vec{\boldsymbol{d}} \otimes \vec{\boldsymbol{d}} \right\rangle = \begin{pmatrix} \boldsymbol{C}^\mathrm{bb} & \boldsymbol{C}^\mathrm{bd}\\ \boldsymbol{C}^\mathrm{db} & \boldsymbol{C}^\mathrm{dd}\end{pmatrix}
\end{equation}
splits into four submatrices for the different auto- and cross-correlations of the two particle species,
\begin{align}
	\boldsymbol{C}^{\alpha\gamma} &\coloneqq \left\langle \boldsymbol{d}^\alpha\otimes\boldsymbol{d}^\gamma\right\rangle = C^{\alpha\gamma}_{jk}\otimes (\vec{e}^\alpha_j\otimes\vec{e}^\gamma_k) \\
	C^{\alpha\gamma}_{jk} &= 
	\begin{pmatrix}
	C^{\alpha\gamma}_{\delta_j\delta_k} & \; \bigl(\vec{C}^{\alpha\gamma}_{\delta_jp_k}\bigr)^\intercal \\
	\vec{C}^{\alpha\gamma}_{p_j\delta_k} & C^{\alpha\gamma}_{p_jp_k}
	\end{pmatrix}
	\coloneqq
	\revision{\begin{pmatrix}
	\langle \delta^{\alpha(\mathrm{i})}_j \delta^{\gamma(\mathrm{i})}_k \rangle& \langle \delta^{\alpha(\mathrm{i})}_j \vec{P}^{\gamma(\mathrm{i})}_k \rangle^\intercal\\
	\langle \vec{P}^{\alpha(\mathrm{i})}_j\delta^{\gamma(\mathrm{i})}_k \rangle \quad & \langle \vec{P}^{\alpha(\mathrm{i})}_j \otimes \vec{P}^{\gamma(\mathrm{i})}_k \rangle
	\end{pmatrix}}.
\end{align}

Due to homogeneity and isotropy of the Gaussian random field, one can draw the following three conclusions for the entries of the covariance matrix:
\begin{itemize}
	\item $C^{\alpha\gamma}_{\delta_j p_k} = 0$ for $\alpha=\gamma$ and $j=k$, i.e.~for the same particle
	\item $C^{\alpha \gamma}_{p_j p_k}$ is always a diagonal matrix
	\item For $\alpha=\gamma$ and $j=k$ both $C^{\alpha\gamma}_{p_jp_k}$ and $C^{\alpha\gamma}_{\delta_j\delta_k}$ must be diagonal matrices which are spatially constant. Specifically, $C^{\alpha\alpha}_{\delta_j\delta_j}=(\sigma^{\alpha}_\delta)^2$ and $C^{\alpha\alpha}_{p_jp_j}=(\sigma^{\alpha}_p)^2\mathbbm
	{1}_3$ with constants $\sigma^{\alpha}_\delta$ and $\sigma_p^{\alpha}$.
\end{itemize}
\revision{If we additionally relate the initial density contrast and momentum fields via the linearised continuity equation, the covariance matrix entries can be expressed solely in terms of the initial density contrast auto- and cross-power spectra $P_\delta^{\alpha \gamma \, (\mathrm{i})}$ of the two particle species,
\begin{eqnarray}
C^{\alpha \gamma}_{\delta_j \delta_k} &=& \int_k P_\delta^{\alpha \gamma \, (\mathrm{i})}(k) \, \mathrm{e}^{-\im \vec{k} \cdot (\vec{q}^{(\mathrm{i})}_j-\vec{q}^{(\mathrm{i})}_k)}\\
 C^{\alpha \gamma}_{\delta_j p_k} &=& \int_k \frac{\vec{k}}{k^2} \, P_\delta^{\alpha \gamma \, (\mathrm{i})}(k) \, \mathrm{e}^{-\im \vec{k} \cdot (\vec{q}^{(\mathrm{i})}_j-\vec{q}^{(\mathrm{i})}_k)} \\
C^{\alpha \gamma}_{p_j p_k} &=& \int_k \frac{\vec{k} \otimes \vec{k}}{k^4} \, P_\delta^{\alpha \gamma \, (\mathrm{i})}(k) \, \mathrm{e}^{-\im \vec{k} \cdot (\vec{q}^{(\mathrm{i})}_j-\vec{q}^{(\mathrm{i})}_k)} \,.
\end{eqnarray}}

Inserting \eqref{Pmicro} and \eqref{Pmacro} into \eqref{Ptotal} and using the properties of the covariance matrix, the expression for the initial phase-space distribution $P(\vec{\boldsymbol{x}}^{(\mathrm{i})})$ can be further simplified. The calculation is completely analogous the case of a system with only one particle species. Hence, we refer  to \cite{bartelmann_microscopic_2016} for a more detailed calculation and directly state the result
\begin{align}
P(\vec{\boldsymbol{x}}^{(\mathrm{i})}) =
\frac{V^{-N}}{\sqrt{(2\pi)^{3N} \det \boldsymbol{C}_{pp}}} \; 
\mathcal{C} \bigg( \frac{\partial}{\mathrm{i}\partial \vec{\boldsymbol{p}}^{(\mathrm{i})}} \bigg) \exp \bigg\{ -\frac{1}{2} \, \vec{\boldsymbol{p}}^{(\mathrm{i})\intercal} \, \boldsymbol{C}^{-1}_{pp} \, \vec{\boldsymbol{p}}^{(\mathrm{i})}  \bigg\} 
\end{align}
with the polynomial
\begin{align}
\mathcal{C} \bigg( \frac{\partial}{\mathrm{i}\partial \vec{\boldsymbol{p}}^{(\mathrm{i})}} \bigg) &\coloneqq \prod_{\mu,\nu\in\{\mathrm{b},\mathrm{d}\}} \, \prod_{n=1}^{N^\mu} \bigg( 1 - \mathrm{i} \sum_{m=1}^{N^\nu} \vec{C}^{\mu\nu}_{\delta_np_m}\cdot \frac{\partial}{\mathrm{i}\partial \vec{p}^{\gamma(\mathrm{i})}_m} \bigg)  \label{polynomial def} \\
&\quad\;+ \sum_{\alpha,\gamma \in\{\mathrm{b},\mathrm{d}\}} \, \sum_{\{i^\alpha,j^\gamma\}} C^{\alpha\gamma}_{\delta_\mathrm{i} \delta_j} \bigg[ \prod_{\mu,\nu} \, \prod_{n=1}^{N^\mu} \bigg( 1- \mathrm{i} \sum_{m=1}^{N^\nu} \vec{C}^{\mu \nu}_{\delta_np_m}\cdot \frac{\partial}{\mathrm{i}\partial \vec{p}^{\nu(\mathrm{i})}_m} \bigg) \bigg]   \nonumber\\
&\quad\;+  \sum_{\alpha,\gamma,\epsilon,\lambda \in\{\mathrm{b},\mathrm{d}\}} \, \sum_{\{\{i^\alpha,j^\gamma\},\{k^\epsilon,\ell^\lambda\}\}} C^{\alpha\gamma}_{\delta_\mathrm{i} \delta_j} C^{\epsilon\lambda}_{\delta_k\delta_\ell} \bigg[ \prod_{\mu,\nu} \, \prod_{n=1}^{N^\mu} \bigg( 1- \mathrm{i} \sum_{m=1}^{N^\nu} \vec{C}^{\mu \nu}_{\delta_np_m}\cdot \frac{\partial}{\mathrm{i}\partial \vec{p}^{\nu(\mathrm{i})}_m} \bigg) \bigg]   \nonumber\\
&\quad\;+ \dots  \nonumber.
\end{align}
Here, the Greek indices denote the particle species while the Latin indices label the individual particles within a species.
There are certain restrictions on the sums and products we have to take care of:
\begin{itemize}
	\item First, in each term a specific particle, i.e.~a specific label combination $j^\alpha$, is allowed to appear only once. For example, in the second line the sum and the products inside the squared brackets exclude those particles which we already sum over outside the brackets. The same holds in the third line. Note that two different particles can carry the same species or number label while they are not identical. For example,~$1^\mathrm{b}$ is the first baryonic particle which is of course not the same as the first DM particle $1^\mathrm{d}$.
	\item  The sum $\sum_{\{i^\alpha,j^\gamma\}}$ runs over all distinct particle pairs. For instance,~$\{1^\mathrm{b},2^\mathrm{d}\}$ is equivalent to $\{2^\mathrm{d},1^\mathrm{b}\}$ while $\{1^\mathrm{b},2^\mathrm{d}\}$ and $\{1^\mathrm{b},2^\mathrm{b}\}$ are not. In addition, it still holds that a specific particle must only appear once. Hence,~pairs of identical particles, like $\{1^\mathrm{b},1^\mathrm{b}\}$, are excluded from the sum. 
	\item The sum $\sum_{\{\{i^\alpha,j^\gamma\},\{k^\epsilon,\ell^\lambda\}\}}$ has to be treated similarly. It runs over all distinct 2-tupels of distinct pairs. For instance, we identify $\{ \{1^\mathrm{b},2^\mathrm{b}\},\{3^\mathrm{b},4^\mathrm{b}\} \}$ with $\{ \{4^\mathrm{b},3^\mathrm{b}\},\{1^\mathrm{b},2^\mathrm{b}\} \}$.
\end{itemize}
This scheme continues to all higher orders.

\section{Derivation of the free generating functional} \label{App: Free Gen}
The calculation of the free generating functional via the expansion in terms of particle clusters is very technical and extensive. Since most of the steps are completely analogous to the case of a single particle species we only give here the main steps showing some central modifications and otherwise refer to the original paper \cite{fabis_kinetic_2018}.

We start by inserting the initial distribution \eqref{MPH: Initial Distribution} into our earlier expression \eqref{Free Generating Functional} for the free generating functional. Splitting the momentum covariance matrix into auto- and cross-correlations according to
\begin{align}
\vec{\boldsymbol{\mathcal{J}}}_{p}^\intercal \boldsymbol{C}_{pp} \vec{\boldsymbol{\mathcal{J}}}_{p} = &\sum_{\alpha,\gamma \in \{\ba,\dm\}} \sum_{j,k=1}^{N^\alpha,N^\gamma} \mathcal{J}^\alpha_{p_j} C^{\alpha\gamma}_{p_jp_k} \mathcal{J}^{\gamma}_{p_k}  \\
= &\sum_{\alpha \in \{\ba,\dm\}} \, \sum_{j=1}^{N^\alpha} \mathcal{J}^{\alpha}_{p_j} (\sigma_p^{\alpha})^2 \mathcal{J}^{\alpha}_{p_j} +
\sum_{\alpha,\gamma \in \{\ba,\dm\}} \sum_{j,k=1}^{N^\alpha,N^\gamma} \mathcal{J}^\alpha_{p_j} C^{\alpha\gamma}_{p_jp_k} \mathcal{J}^{\gamma}_{p_k} \, \biggr|_{j^\alpha \neq k^\gamma}  \nonumber\\
\eqqcolon &\sum_{\alpha \in \{\ba,\dm\}} \, \sum_{j=1}^{N^\alpha} \mathcal{J}^{\alpha}_{p_j} (\sigma_p^{\alpha})^2 \mathcal{J}^{\alpha}_{p_j} +
\vec{\boldsymbol{\mathcal{J}}}_{p}^\intercal \tilde{\boldsymbol{C}}_{pp} \vec{\boldsymbol{\mathcal{J}}}_{p},    \nonumber
\end{align}
with 
\begin{align}
	&\boldsymbol{\mathcal{J}}^\mathrm{\alpha}_{q/p} \coloneqq \int_{t_\mathrm{i}}^\infty \mathrm{d}t\ \boldsymbol{J}^{\alpha \, \intercal}(t) \, \boldsymbol{\mathcal{G}}^{\mathrm{R} \alpha}(t,t_\mathrm{i}) \, \boldsymbol{P}^{\alpha}_{q/p} \eqqcolon \mathcal{J}^\alpha_{q_j/p_j} \otimes \vec{e}^\alpha_j , \\
	&\boldsymbol{P}_q^{\alpha} \coloneqq 
	\begin{pmatrix}
	\mathbbm{1}_3 \\ 0_3
	\end{pmatrix}
	\otimes
	\mathbbm{1}_{N^\alpha} ,
	\qquad
	\boldsymbol{P}_p^{\alpha} \coloneqq 
	\begin{pmatrix}
	0_3 \\ \mathbbm{1}_3
	\end{pmatrix}
	\otimes
	\mathbbm{1}_{N^\alpha} ,
\end{align}
where $0_3\coloneqq (0,0,0)^\intercal$, we find
\begin{align}
Z_{0}[\vec{\boldsymbol{J}},\vec{\boldsymbol{K}}] = \int \mathrm{d}\vec{\boldsymbol{x}}^{(\mathrm{i})}\  &\hat{\mathcal{C}}_\mathrm{tot}\bigg( \frac{\updelta}{\mathrm{i} \updelta \vec{\boldsymbol{K}}_p(t_\mathrm{i})} \bigg) \, \frac{P_{\sigma^{\mathrm{b}}_p}\bigl(\boldsymbol{p}^{\mathrm{b}(\mathrm{i})}\bigr)}{V^{N^\mathrm{b}}} \, \frac{P_{\sigma^{\mathrm{d}}_p}\bigl(\boldsymbol{p}^{\mathrm{b}(\mathrm{i})}\bigr)}{V^{N^\mathrm{d}}}    \label{MPH: Appendix Coupled Free Gen Func}\\ 
&\times \exp\bigg\{ \mathrm{i} \bigg( \vec{\boldsymbol{\mathcal{J}}}_q \cdot \vec{\boldsymbol{q}}^{(\mathrm{i})} + \vec{\boldsymbol{\mathcal{J}}}_p \cdot \vec{\boldsymbol{p}}^{(\mathrm{i})} -S_K[\vec{\boldsymbol{J}},\vec{\boldsymbol{K}}] \bigg) \bigg\}  \nonumber,
\end{align}
where
\begin{align}
	S_K[\vec{\boldsymbol{J}},\vec{\boldsymbol{K}}] &\coloneqq \sum_{\alpha \in \{\ba,\dm\}} \int \mathrm{d}t \, \mathrm{d}t' \, \boldsymbol{J}^{\alpha \, \intercal}(t) \boldsymbol{\mathcal{G}}^\mathrm{R \alpha}(t,t') \boldsymbol{K}^\alpha(t'), \\
	P_{\sigma^{\alpha}_p} (\boldsymbol{p}^{\alpha(\mathrm{i})}) &\coloneqq \frac{1}{\big(2\pi(\sigma^{\alpha }_p)^2\big)^{3N^\alpha/2}} \exp \bigg\{ -\frac{\boldsymbol{p}^{\alpha(\mathrm{i})\intercal}\boldsymbol{p}^{\alpha (\mathrm{i})}}{2 (\sigma^{\alpha}_p)^2} \bigg\}   \label{Gaussian distr}
\end{align}
and
\begin{align}
\hat{\mathcal{C}}_\mathrm{tot} \bigg( \frac{\updelta}{\mathrm{i} \updelta \vec{\boldsymbol{K}}_p(t_\mathrm{i})} \bigg) \coloneqq \hat{\mathcal{C}} \bigg( \frac{\updelta}{\mathrm{i} \updelta \vec{\boldsymbol{K}}_p(t_\mathrm{i})} \bigg) \exp \bigg\{ -\frac{1}{2} \bigg( \frac{\updelta}{\mathrm{i} \updelta \vec{\boldsymbol{K}}_p(t_\mathrm{i})} \bigg)^\intercal \tilde{\boldsymbol{C}}_{pp} \bigg( \frac{\updelta}{\mathrm{i} \updelta \vec{\boldsymbol{K}}_p(t_\mathrm{i})} \bigg) \bigg\}.
\end{align}
As mentioned in \autoref{Sec: Free GenFunc}, for vanishing initial cross-correlation (corresponding to $\hat{\mathcal{C}}_\mathrm{tot} \rightarrow 1$) the free generating functional factorizes into single-particle contributions. Considering non-vanishing cross-correlations, the particles become connected and the resulting factorization can now be performed in terms of clusters of correlated particles.

An important relation found according to this factorization is that the free generating functional can be written as
\begin{equation}
Z_{0}[\vec{\boldsymbol{J}},\vec{\boldsymbol{K}}] 
= N^\mathrm{b}! N^\mathrm{d}! \sum_{\{m_{\ell^\mathrm{b}\ell^\mathrm{d}}\}^*} \prod_{\substack{\ell^\mathrm{b},\ell^\mathrm{d} = 0 \\ \ell^\mathrm{b}+\ell^\mathrm{d} \geq 1}}^{N^\mathrm{b},N^\mathrm{d}} \frac{\big( W_0^{(\ell^\mathrm{b},\ell^\mathrm{d})} [\vec{\boldsymbol{J}},\vec{\boldsymbol{K}}]\big)^{m_{\ell^\mathrm{b}\ell^\mathrm{d}}}}{m_{\ell^\mathrm{b}\ell^\mathrm{d}}!},   \label{MPH: Expansion in W of Free Can Gen Func}
\end{equation}
where $W^{(\ell^\mathrm{b},\ell^\mathrm{d})}_0 $ denotes the free generating functional of cumulant contributions from clusters of exactly $\ell^\mathrm{b}$ correlated baryonic and $\ell^\mathrm{d}$ correlated dark-matter particles. The sum in \eqref{MPH: Expansion in W of Free Can Gen Func} runs over all possible ways to distribute the $N^\mathrm{b}+N^\mathrm{d}$ particles over a collection of such clusters, where $m_{\ell^\mathrm{b} \ell^\mathrm{d}}$ is the number of $(\ell^\mathrm{b},\ell^\mathrm{d})$-particle clusters appearing in each collection. Thereby, each possible collection needs to satisfy the constraints $\sum_{\ell^\mathrm{b}=1}^{N^\mathrm{b}} \sum_{\ell^\mathrm{d}=1}^{N^\mathrm{d}} \ell^\alpha m_{\ell^\mathrm{b} \ell^\mathrm{d}} = N^\alpha$ for $\alpha \in \{\mathrm{b},\mathrm{d}\}$.

To compute the $W^{(\ell^\mathrm{b},\ell^\mathrm{d})}_0$ systematically, \cite{fabis_kinetic_2018} introduced a diagrammatic representation for the different ways particles can be correlated with each other within a cluster. These representations can be easily generalised for the case of two particle species by introducing different types of diagrams for the different particle species. The main difference between one- and two-species systems is then purely combinatorial: For two distinguishable species of particles, there are more distinct possibilities to distribute the particles among clusters.

\section{Derivation of the free cumulants} \label{App: Free Cumulants}
As for the derivation of the free generating functional, the derivation of the free collective-field cumulants is very technical and extensive. Hence, we give only some crucial steps while referring to \cite{fabis_kinetic_2018} for more details.

To calculate the free cumulants it is convenient to work in a grand canonical ensemble, where the particle number is not fixed anymore. The reason for this is that the weighted sum over different particle numbers involved in this transition simplifies the form of the generating functional \eqref{MPH: Expansion in W of Free Can Gen Func} significantly,
\begin{equation}
	Z_{0}[\vec{\boldsymbol{J}},\vec{\boldsymbol{K}}] \rightarrow
 \exp\left\{\vphantom{\sum_{\ell^\mathrm{b} = 0}^\infty}\right. \sum_{\substack{\ell^\mathrm{b},\ell^\mathrm{d} = 0 \\ \ell^\mathrm{b}+\ell^\mathrm{d} \geq 1}}^\infty W_0^{(\ell^\mathrm{b},\ell^\mathrm{d})} [\vec{\boldsymbol{J}},\vec{\boldsymbol{K}}]\left.\vphantom{\sum_{\ell^\mathrm{b} = 0}^\infty}\right\} .   \label{MPH: Expansion in W of Free Grand Can Gen Func}
\end{equation}
Note that we can safely make this transition since the canonical and grand canonical ensembles are equivalent in the thermodynamic limit that we are considering.

Analogously to the interacting collective-field cumulants, the free cumulants are obtained by applying collective-field operators to $\ln Z_0$ and setting the source fields to zero afterwards,
\begin{align}
	G^{(0) \, \alpha_1 \dots \alpha_{l_n} \, \gamma_1 \dots \gamma_{l_B}}_{\hphantom{(0)} \, n \cdots n \, B \cdots B}(1,\dots,l_n,1',\dots,l_B') &= \prod_{u=1}^{l_n} \Bigl(\hat{\Phi}^{\alpha_u}_n(u)\Bigr) \, \prod_{r=1}^{l_B} \Bigl(\hat{\Phi}^{\gamma_r}_{B}(r')\Bigr) \sum_{\substack{\ell^\mathrm{b},\ell^\mathrm{d} = 0 \\ \ell^\mathrm{b}+\ell^\mathrm{d} \geq 1}}^\infty W^{(\ell^\mathrm{b},\ell^\mathrm{d})}_0 [\vec{\boldsymbol{J}},\vec{\boldsymbol{K}}] \biggr|_{\vec{\boldsymbol{J}},\vec{\boldsymbol{K}}=0} \nonumber \label{MPH: Coupled Free Cumulants Equation Appendix}\\
&\eqqcolon \sum_{\substack{\ell^\mathrm{b},\ell^\mathrm{d} = 0 \\ \ell^\mathrm{b}+\ell^\mathrm{d} \geq 1}}^\infty G^{(\ell^\mathrm{b},\ell^\mathrm{d}) \, \alpha_1 \dots \alpha_{l_n} \, \gamma_1 \dots \gamma_{l_B}}_{0 \hphantom{(,\ell^\mathrm{d})} \;\, n \cdots n \, B \cdots B} (1,\dots,l_n,1',\dots,l_B') .
\end{align}
Here, $\alpha_u, \gamma_r \in \{\mathrm{b},\mathrm{d} \}$ label to which particle species the $u$-th density field and the $r$-th response field correspond, respectively.

As with the free generating functional, the explicit computation of the free cumulants proceeds completely analogously to the case of a single particle species detailed in \cite{fabis_kinetic_2018}, with the main difference being a higher number of different combinatorial contributions that have to be considered for the possible ways particles of two distinct species can be correlated with each other. This leads to the following rules a general $(\ell^\mathrm{b},\ell^\mathrm{d})$-cumulant needs to obey, which are the natural generalisations of the rules found in \cite{fabis_kinetic_2018} for the single-species case:
\begin{enumerate}
	\item \label{homogeneity rule} In a statistically homogeneous system, an $(\ell^\mathrm{b},\ell^\mathrm{d})$-cumulant vanishes if for any $\alpha \in \{\mathrm{b},\mathrm{d}\}$ the number of $\alpha$-particles $\ell^\alpha$ is larger than the number of $\Phi_n^\alpha$-fields appearing in the cumulant,
	\begin{align}
		&G^{(\ell^\mathrm{b},\ell^\mathrm{d}) \, \alpha_1\dots \alpha_{l_n} \gamma_1 \dots \gamma_{l_B}}_{0 \qquad\; n \cdots n B \cdots B}(1, \dotsc, l_n, 1', \dotsc l'_B) = 0 \quad \text{if} \quad \exists \, \alpha \in \{\mathrm{b},\mathrm{d}\} \\
		&\quad \text{such that} \quad \ell^\alpha > \bigl|\bigl\{i \in \{1,\dotsc,l_n\} \, | \, \alpha_i = \alpha \bigr\}\bigr| , \nonumber
	\end{align}
	were $|\{\dots\}|$ denotes the number of elements of a set.
	\item \label{causality rule} An $(\ell^\mathrm{b},\ell^\mathrm{d})$-cumulant vanishes if for any $\alpha \in \{\mathrm{b},\mathrm{d}\}$ there is a $\Phi_B^\alpha$-field argument evaluated at an equal or later time than all $\Phi_n^\alpha$-field arguments,
	\begin{align}
		&G^{(\ell^\mathrm{b},\ell^\mathrm{d}) \, \alpha_1\dots \alpha_{l_n} \gamma_1 \dots \gamma_{l_B}}_{0 \qquad n \cdots n B \cdots B}(1, \dotsc, l_n, 1', \dotsc l'_B) = 0 \quad\; \text{if} \quad \exists \, i \in \{1,\dotsc,l_B\} \\
		&\quad\text{such that} \quad t_i \geq t_j \quad \forall \, j \in \{1,\dotsc,l_n \, | \, \alpha_j = \gamma_i\} . \nonumber
	\end{align}
\end{enumerate}
From rule 1 we can draw the important conclusion that the sums over $\ell^\alpha$ in \eqref{MPH: Coupled Free Cumulants Equation Appendix} truncate at the respective numbers of density fields $\Phi_n^\alpha$ appearing in the cumulant. Rule 2 implies that any free collective-field cumulant vanishes if it involves response fields $\Phi_B^\alpha$ but no density fields $\Phi_n^\alpha$ for any of the particle species.

For our later purposes, we are interested in the 2-point cumulants. According to rule 1 the only non-vanishing and non-negligible contributions are coming from terms with exactly two particles, i.e.~$(\ell^\mathrm{b}=2,\ell^\mathrm{d}=0)$, $(\ell^\mathrm{b}=1,\ell^\mathrm{d}=1)$ or $(\ell^\mathrm{b}=0,\ell^\mathrm{d}=2)$, since the terms corresponding to only one particle give shot-noise contributions and can thus be neglected because of the large number of particles in cosmologically relevant volumes. Due to rule 2, the pure $\vec{\Phi}_B$-field cumulant vanishes, i.e.
\begin{equation}
G^{(0) \alpha_1\alpha_2}_{\ \ \  BB}(1,2)=0 \qquad \forall \, \alpha_1,\alpha_2 \in \{\mathrm{b},\mathrm{d}\} .
\end{equation}
Similarly, the mixed $\Phi_n^{\alpha_1}$- and $\Phi_B^{\alpha_2}$-field cumulant vanishes if the two fields correspond to different particle species, $\alpha_1 \neq \alpha_2$, leaving us with
\begin{align}
G^{(0) \alpha_1 \alpha_2}_{\ \ \ nB}(1,2) &= G^{(0) \alpha_2 \alpha_1}_{\ \ \ Bn}(2,1)
= \updelta_{\alpha_1 \alpha_2} \, \bar{n}^{\alpha_1} (2\pi)^3 \, \updelta_\textsc{d}\bigl(\vec{L}_{q,1} + \vec{L}_{q,2}\bigr) \\
&\quad\;\times \Big(\mathrm{i} \vec{k}_2 \cdot \vec{L}_{p,1}(t_2) \Big) \, \exp \bigg\{ - \frac{(\sigma^{\alpha_1}_p)^2}{2} (\vec{L}_{p,1} + \vec{L}_{p,2})^2 \bigg\}. \nonumber
\end{align}
Here, the spatial shift vectors
\begin{align}
\vec{L}_{q,r}(t) &\coloneqq \vec{k}_r g_{qq}(t_r,t) , \quad \vec{L}_{q,r} \coloneqq \vec{L}_{q,r}(t_\mathrm{i})   \\
\vec{L}_{p,r}(t) &\coloneqq \vec{k}_r g_{qp}(t_r,t) ,  \quad \vec{L}_{p,r} \coloneqq \vec{L}_{p,r}(t_\mathrm{i})
\end{align}
encode how the phase of the Fourier transformed density is changed by the free motion of particles from time $t$ to $t_r$. For the pure $\vec{\Phi}_n$-field cumulant we obtain at linear order in the initial density contrast power spectra $P^{\alpha \gamma \, \mathrm{(i)}}_\delta$ of baryonic and dark matter 
\begin{align}
G^{(0) \alpha_1\alpha_2}_{\ \ \ nn}(1,2)\Bigr|_\mathrm{lin} &= \, (\bar{n}^\mathrm{b})^{\ell^\mathrm{b}} (\bar{n}^\mathrm{d})^{\ell^\mathrm{d}} (2\pi)^3 \, \updelta_\textsc{d}\bigl(\vec{L}_{q,1} + \vec{L}_{q,2}\bigr) \, 
P^{\alpha \gamma \, \mathrm{(i)}}_\delta(k_1) \nonumber \\
&\quad\;\times \bigg( 1 + \frac{\vec{L}_{p,1} \cdot\vec{k_1}}{k_1^2} \bigg) \bigg( 1 + \frac{\vec{L}_{p,2} \cdot \vec{k_2}}{k_2^2} \bigg) \\
&\quad\;\times \exp \bigg\{ - \frac{(\sigma^{\alpha_1}_p)^2}{2} \vec{L}^2_{p,1}  - \frac{(\sigma^{\alpha_2}_p)^2}{2} \vec{L}^2_{p,2} \bigg\} . \nonumber
\end{align}

Note that \cite{fabis_kinetic_2018} uses the Klimontovich phase-space density field and the respective phase-space response field instead of the spatial density and response fields used in this work. However, the generalization to phase-space fields is straightforward, as the form of the free collective-field cumulants remains unchanged and only the phase-shift vectors $\vec{L}_{q,r}$ and $\vec{L}_{p,r}$ need to be replaced by their respective phase-space equivalents.

\section{\revision{Computation of the linearly evolved power spectra}} \label{App:linear_power_spectra_computation}
\revision{To compute the linearly evolved dark, baryonic and total matter power spectra from tree-level RKFT, we first exploit the statistical homogeneity of cosmic structure formation, allowing us to express the macroscopic propagator components and the free cumulants as
\begin{align}
    \tilde{\Delta}_{\vec{n}\vec{n}}(1, 2) \eqqcolon &\,(2\pi)^3 \, \delta_\mathrm{D}(\vec{k}_1 + \vec{k}_2) \, \tilde{\Delta}_{\vec{n} \vec{n}}(k_1; \eta_1, \eta_2) \,, \label{statistical_propagator_in_homogeneous_system} \\
    \tilde{\Delta}_\textsc{r}(1, 2) = \tilde{\Delta}_\textsc{a}(2, 1) \eqqcolon &\,(2\pi)^3 \, \delta_\mathrm{D}(\vec{k}_1 + \vec{k}_2) \, \tilde{\Delta}_\textsc{r}(k_1; \eta_1, \eta_2) \label{causal_propagator_in_homogeneous_system} \\
    =\, &\,(2\pi)^3 \, \delta_\mathrm{D}(\vec{k}_1 + \vec{k}_2) \, \tilde{\Delta}_\textsc{a}(k_1; \eta_2, \eta_1)\,, \nonumber \\
    G^{(0)}_{\vec{n} \vec{n}}(1, 2) \eqqcolon &\,(2\pi)^3 \, \delta_\mathrm{D}(\vec{k}_1 + \vec{k}_2) \, G^{(0)}_{\vec{n} \vec{n} \underline{v}}(k_1; \eta_1, \eta_2) \,, \label{free_nn_cumulant_in_homogeneous_system} \\
    G^{(0)}_{\vec{n} \vec{B}}(1, 2) \, \underline{v}(2) \eqqcolon &\,(2\pi)^3 \, \delta_\mathrm{D}(\vec{k}_1 + \vec{k}_2) \, G^{(0)}_{\vec{n} \vec{B} \underline{v}}(k_1; \eta_1, \eta_2) \,. \label{free_nBv_cumulant_in_homogeneous_system}
\end{align}
The next step is to solve the integral equation \eqref{integral equation for retarded propagator} for the retarded and advanced macroscopic propagators $\tilde{\Delta}_\textsc{r}$ and $\tilde{\Delta}_\textsc{a}$. Using \eqref{causal_propagator_in_homogeneous_system} and \eqref{free_nBv_cumulant_in_homogeneous_system}, this simplifies to an integral equation in time only,
\begin{equation}
\tilde{\Delta}_\textsc{r}(k_1; \eta_1, \eta_2) = - \mathrm{i} G^{(0)}_{\vec{n} \vec{B} \underline{v}}(k_1; \eta_1, \eta_2) - \int_{\eta_2}^{\eta_1} \mathrm{d}\bar{\eta} \; \mathrm{i} G^{(0)}_{\vec{n} \vec{B} \underline{v}}(k_1; \eta_1, \bar{\eta}) \, \tilde{\Delta}_\textsc{r}(k_1; \bar{\eta}, \eta_2) \,.
\label{retarded_propagator_simplified_integral_equation}
\end{equation}
The function $G^{(0)}_{\vec{n} \vec{B} \underline{v}}$ is obtained by inserting the expressions for the interaction potentials \eqref{dm potential} and \eqref{b potential} in the limit of ideal hydrodynamics, $\sigma_0 \rightarrow 0$, as well as the free $\vec{n}\vec{B}$-cumulant \eqref{2-pt Cumulants 2}. Together with the scale function \eqref{scalefunction} and the relations
\eqref{mean_density_relation}, \eqref{c_g with total density} and \eqref{c_p specialised}, this yields
\begin{equation}
    - \mathrm{i} G^{(0)}_{\vec{n} \vec{B} \underline{v}}(k_1; \eta_1, \eta_2) = \frac{3}{2} \, \frac{g_{qp}(\eta_1,\eta_2) \, H_0^2}{a_\im^2 \, a^2(\eta_2) \, H_\im \, H(\eta_2)} \,
    \begin{pmatrix}
        \Omega_{\mathrm{m},0}^\ba - \frac{10}{9} \, k_1^2 \, \frac{k_\mathrm{B} \, T(\eta_2)}{m_\mathrm{P} H_0^2} & \quad \Omega_{\mathrm{m},0}^\ba \\
        \Omega_{\mathrm{m},0}^\dm & \quad \Omega_{\mathrm{m},0}^\dm
    \end{pmatrix} \,,
    \label{free_nBv_cumulant_cosmo}
\end{equation}
where $a_\im$ and $H_\im$ denote the initial value of the scale factor and the Hubble function, respectively. The time-dependence of the mean gas temperature $T$ is specified in \eqref{temperature evolution}, and the $qp$-propagator is given by
\begin{equation}
    g_{qp}(\eta_1,\eta_2) = \Theta(\eta_1,\eta_2) \, \int_{\eta_2}^{\eta_1} \, \mathrm{d} \bar{\eta} \, \frac{a_\im^2 \, H_\im}{a^2(\bar{\eta}) \, H(\bar{\eta})} \,.
    \label{qp_propagator_cosmo}
\end{equation}
Once the cosmology and thus the evolution of Hubble function $H(\eta)$ and scale factor $a(\eta)$ are fixed, the integral equation \eqref{retarded_propagator_simplified_integral_equation} can be approximated as a linear $2N_t \times 2N_t$ matrix equation, by approximating the time integral as a direct sum over $N_t$ time steps. Due to the causal structure of \eqref{qp_propagator_cosmo}, the matrix-equivalent of $G^{(0)}_{\vec{n} \vec{B} \underline{v}}$ is of lower block-triangular form, which renders solving this equation computationally inexpensive.}

\revision{Afterwards, we use the solution for $\tilde{\Delta}_\textsc{r}$ to compute the $\Delta_{\vec{n}\vec{n}}$-component of the macroscopic propagator according to \eqref{nn-component of the propagator}. Using \eqref{statistical_propagator_in_homogeneous_system} to \eqref{free_nn_cumulant_in_homogeneous_system}, this simplifies to performing 1- and 2-dimensional time integrals,
\begin{align}
\Delta_{\vec{n}\vec{n}}(k_1;\eta_1,\eta_2) = G^{(0)}_{\vec{n}\vec{n}}(k_1;\eta_1,\eta_2)
&+ \int_0^{\eta_1} \mathrm{d}\bar{\eta}_1 \; \tilde{\Delta}_\textsc{r}(k_1;\eta_1,\bar{\eta}_1) \, G^{(0)}_{\vec{n}\vec{n}}(k_1;\bar{\eta}_1,\eta_2) \nonumber \\
&+ \int_0^{\eta_2} \mathrm{d}\bar{\eta}_2 \; G^{(0)}_{\vec{n}\vec{n}}(k_1;\eta_1,\bar{\eta}_2) \, \tilde{\Delta}_\textsc{a}(k_1;\bar{\eta}_2,\eta_2) \label{statistical_propagator} \\
&+ \int_0^{\eta_1} \mathrm{d}\bar{\eta}_1 \int_0^{\eta_2} \mathrm{d}\bar{\eta}_2 \; \tilde{\Delta}_\textsc{r}(k_1;\eta_1,\bar{\eta}_1) \, G^{(0)}_{\vec{n}\vec{n}}(k_1;\bar{\eta}_1,\bar{\eta}_2) \, \tilde{\Delta}_\textsc{a}(k_1;\bar{\eta}_2,\eta_2) \,, \nonumber
\end{align}
where we use the relation \eqref{mean_density_relation} to express the free 2-point density cumulant \eqref{2-pt Cumulants 3} as
\begin{align}
G^{(0)}_{\vec{n}\vec{n}}(k_1;\eta_1,\eta_2) &= \biggl(\frac{\bar{n}}{\Omega_{\mathrm{m},0}}\biggr)^2 \, \big(1+g_{qp}(\eta_1,0)\big)\big(1+g_{qp}(\eta_2,0)\big) \label{free_nn_cumulant_cosmo} \\
&\quad\;\times\begin{pmatrix}
\Omega_{\mathrm{m},0}^\ba \Omega_{\mathrm{m},0}^\ba \, P^{\mathrm{bb} \, (\mathrm{i})}_\delta(k_1) \quad & \Omega_{\mathrm{m},0}^\ba \Omega_{\mathrm{m},0}^\dm \, P^{\mathrm{bd} \, (\mathrm{i})}_\delta(k_1)\\
\Omega_{\mathrm{m},0}^\dm \Omega_{\mathrm{m},0}^\ba \, P^{\mathrm{db} \, (\mathrm{i})}_\delta(k_1)\quad & \Omega_{\mathrm{m},0}^\dm \Omega_{\mathrm{m},0}^\dm \, P^{\mathrm{dd} \, (\mathrm{i})}_\delta(k_1) \\
\end{pmatrix} \,. \nonumber
\end{align}
Finally, according to \eqref{individual_particle_species_power_spectrum} and \eqref{total_matter_power_spectrum}, the linear tree-level power spectra of dark, baryonic and total matter are obtained from the equal-time components of \eqref{statistical_propagator},
\begin{align}
P_\delta^{\alpha \gamma \, (\mathrm{tree})}(k_1,\eta_1) &= \frac{1}{\bar{n}^2} \, \frac{(\Omega_{\mathrm{m},0})^2}{\Omega_{\mathrm{m},0}^\alpha \, \Omega_{\mathrm{m},0}^\gamma} \, \Delta^\mathrm{\alpha \gamma}_{nn}(k_1; \eta_1, \eta_1) \\ 
P_\delta^\mathrm{(tot, tree)}(k_1,\eta_1) &= \frac{1}{\bar{n}^2} \, \Big( \Delta^\mathrm{bb}_{nn}(k_1; \eta_1, \eta_1) + \Delta^\mathrm{bd}_{nn}(k_1; \eta_1, \eta_1) \\
& + \Delta^\mathrm{db}_{nn}(k_1; \eta_1, \eta_1) + \Delta^\mathrm{dd}_{nn}(k_1; \eta_1, \eta_1) \Big) \,. \nonumber 
\end{align}
Note that the total comoving mean number density $\bar{n}$ cancels out when inserting \eqref{statistical_propagator} and \eqref{free_nn_cumulant_cosmo}. We can thus simply set $\bar{n} = 1$ for this calculation.}

\bibliography{bibliography}

\end{document}